\documentclass[aps,prd,amsmath,floats,floatfix, twocolumn,
superscriptaddress,nofootinbib,showpacs]{revtex4-1}

\newcommand{\icecube}[0]{IceCube}

\def\newacronym#1#2#3{\gdef#1{#3 (#2)\gdef#1{#2}}}

\def\prd{Physical Review D }       
\def\prl{Physical Review Letters }    

\def\physrep{Physics Reports} 
\def\araa{Annual Reviews of Astronomy \& Astrophysics}

\newacronym{\gt}{Georgia Tech}{Georgia Institute of Technology}
\newacronym{\icecube}{IceCube}{IceCube South Pole Neutrino Observatory}
\newacronym{\cta}{CTA}{Cherenkov Telescope Array}
\newacronym{\sdsc}{SDSC}{San Diego Supercomputer Center}
\newacronym{\cra}{CRA}{Center for Relativistic Astrophysics}
\newacronym{\nr}{NR}{numerical relativity}
\newacronym{\ornl}{ORNL}{Oak Ridge National Laboratory}
\newacronym{\lisa}{LISA}{Laser Interferometer Space Antenna}
\newacronym{\hst}{HST}{Hubble Space Telescope}
\newacronym{\jwst}{JWST}{James Webb Space Telescope}
\newacronym{\alma}{ALMA}{Atacama Large Millimeter Array}
\newacronym{\ligo}{LIGO}{Laser Interferometer Gravitational Wave Observatory}
\newacronym{\sph}{SPH}{smooth particle hydrodynamics}
\newacronym{\tsi}{TSI}{Terascale Supernova Initiative}
\newacronym{\wmap}{WMAP}{the Wilkinson Microwave Anisotropy Probe}
\newacronym{\cmb}{CMB}{cosmic microwave background}
\newacronym{\ibbh}{IBBH}{intermediate binary black hole}
\newacronym{\grb}{GRB}{Gamma-ray burst}
\newacronym{\grbs}{GRBs}{Gamma-ray bursts}
\newacronym{\hpc}{HPC}{High Performance Computing}
\newacronym{\htc}{HTC}{High Throughput Computing}
\newacronym{\bssn}{BSSN}{Baumgarte-Shapiro-Shibata-Nakamura}
\newacronym{\amr}{AMR}{adaptive mesh refinement}
\newacronym{\mpi}{MPI}{Message Passing Interface}
\newacronym{\cuda}{CUDA}{Compute Unified Device Architecture}
\newacronym{\gui}{GUI}{Graphical User Inferface}
\newacronym{\oit}{OIT}{Office of Information Technologies}
\newacronym{\lbg}{LBGs}{Lyman break galaxies}
\newacronym{\mhd}{MHD}{magneto-hydrodynamics}
\newacronym{\imf}{IMF}{initial mass function}
\newacronym{\sms}{SMS}{supermassive star}
\newacronym{\posix}{POSIX}{Portable Operating System Interface}
\newacronym{\ssd}{SSD}{Solid State Drive}
\newacronym{\os}{OS}{Operating System}
\newacronym{\pace}{PACE}{Partnership for an Advanced Computing Environment}
\newacronym{\oit}{OIT}{Office of Information Technology}
\newacronym{\gtswd}{GTSWD}{Georgia Tech Software Distribution}
\newacronym{\uga}{UGA}{University of Georgia}
\newacronym{\mahpic}{MaHPIC}{Malaria Host-Pathogen Interaction Center}
\newacronym{\nih}{NIH}{National Institutes of Health}
\newacronym{\cdc}{CDC}{Centers for Disease Control and Prevention}
\newacronym{\xsede}{XSEDE}{Extreme Science and Engineering Discovery Environment}
\newacronym{\osg}{OSG}{Open Science Grid}
\newacronym{\egi}{EGI}{European Grid Infrastructure}
\newacronym{\vo} {VO}{Virtual Organization}
\newacronym{\sox}{SoX}{Southern Crossroads}
\newacronym{\slr}{SLR}{Southern Light Rail}
\newacronym{\ddn}{DDN}{DataDirect Networks}
\newacronym{\rsv}{RSV}{Resource and Service Validation}
\newacronym{\cvmfs}{CVMFS}{CernVM File System}
\newacronym{\nfs}{NFS}{Network File Sytem}
\newacronym{\lsc}{LSC}{LIGO Scientific Collaboration}

\newacronym{\DA}{DA}{data analysis}
\newacronym{\CBC}{CBC}{Compact Binary Coalescences}
\newacronym{\lvc}{LVC}{LIGO-Virgo Collaboration}
\newacronym{\aligo}{aLIGO}{Advanced LIGO}
\newacronym{\ldr}{LDR}{LIGO Data Replicator}

\def\si#1{Senior Investigator#1 (SI#1)\gdef\si{SI}}
\def\emri#1{Extreme Mass-Ratio Inspiral#1 (EMRI#1)\gdef\emri{EMRI}}
\def\imbh#1{intermediate mass black hole#1 (IMBH#1)\gdef\imbh{IMBH}}
\def\smbh#1{supermassive black hole#1 (SMBH#1)\gdef\smbh{SMBH}}

\def\bbh#1{binary black hole#1 (BBH#1)\gdef\bbh{BBH}}
\def\bh#1{black hole#1 (BH#1)\gdef\bh{BH}}
\def\ns#1{neutron star#1 (NS#1)\gdef\ns{NS}}
\def\bns#1{binary neutron star#1 (BNS#1)\gdef\bns{BNS}}
\def\nsbh#1{neutron star--black hole#1 (NSBH#1)\gdef\nsbh{NSBH}}
\def\hmns#1{hypermassive neutron star#1 (HMNS#1)\gdef\hmns{HMNS}}
\def\gw#1{gravitational wave#1 (GW#1)\gdef\gw{GW}}
\def\grb#1{Gamma-ray burst#1 (GRB#1)\gdef\grb{GRB}}
\def\pnw#1{post-Newtonian#1 (PN#1)\gdef\pnw{PN}}
\def\eos#1{equation of state#1 (EOS#1)\gdef\eos{EOS}}
\def\gpu#1{graphics processing unit#1 (GPU#1)\gdef\gpu{GPU}}
\def\sn#1{supernova#1 (SN#1)\gdef\sn{SN}}
\def\lae#1{Lyman Alpha Emitter#1 (LAE#1)\gdef\lae{LAE}}
\def\sfr#1{star formation rate#1 (SFR#1)\gdef\sfr{SFR}}
\def\sed#1{spectral energy density#1 (SED#1)\gdef\sed{SED}}
\def\lsst#1{Large Synaptic Survey Telescope#1 (LSST#1)\gdef\lsst{LSST}}
\def\rproc#1{rapid neutron capture#1 ($r$-process#1)\gdef\rproc{$r$-process#1}}

\newacronym{\grhd}{GRHD}{general relativistic hydrodynamics}
\newacronym{\grmhd}{GRMHD}{general relativistic magnetohydrodynamics}
\newacronym{\gw}{GW}{gravitational wave}
\newacronym{\EM}{EM}{electromagnetic}
\newacronym{\utb}{UTB}{University of Texas (Brownsville)}
\newacronym{\aps}{APS}{American Physical Society}
\newacronym{\lsu}{LSU}{Louisiana State University}
\newacronym{\rit}{RIT}{Rochester Institute of Technology}

\usepackage[colorlinks, pdfborder={0 0 0}, plainpages=false]{hyperref}
\usepackage{graphicx}
\usepackage{xspace}
\usepackage[usenames,dvipsnames]{color}
\usepackage{amssymb}
\usepackage[normalem]{ulem} 

\newcommand{\peak}{{\mbox{\tiny peak}}}

\newcommand{\nn}{\nonumber}

\newcommand{\CITA}{\affiliation{Canadian Institute for Theoretical
    Astrophysics, 60 St.~George Street, University of Toronto,
    Toronto, ON M5S 3H8, Canada}} %
\newcommand{\GATech}{\affiliation{Center for Relativistic Astrophysics, School of Physics, Georgia Institute of Technology, Atlanta, Georgia 30332, USA}} %
\newcommand{\HITS}{\affiliation{Heidelberger Institut f\"ur Theoretische Studien, Schloss-Wolfsbrunnenweg 35, D-69118 Heidelberg, Germany}}
\newcommand{\MSU}{\affiliation{eXtreme Gravity Institute, Department of Physics, Montana State University,
Bozeman, MT 59717, USA}}
\newcommand{\NASA}{\affiliation{NASA Marshall Space Flight Center, Huntsville AL 35812, USA}}

\begin{document}

\title{Inferring the post-merger gravitational wave emission from binary neutron star coalescences}

\author{Katerina Chatziioannou}  \CITA
\author{James Alexander Clark} \GATech
\author{Andreas Bauswein} \HITS
\author{Margaret Millhouse}\MSU
\author{Tyson B. Littenberg}\NASA
\author{Neil Cornish}\MSU

\date{\today}

\begin{abstract}
We present a robust method to characterize the gravitational wave emission from the remnant of a neutron star coalescence. Our approach makes only minimal assumptions about the morphology of the signal and provides a full posterior probability distribution of the underlying waveform. We apply our method on simulated data from a network of advanced ground-based detectors and demonstrate the gravitational wave signal reconstruction. We study the reconstruction quality for different binary configurations and equations of state for the colliding neutron stars. We show how our method can be used to constrain the yet-uncertain equation of state of neutron star matter. The constraints on the equation of state we derive are complementary to measurements of the tidal deformation of the colliding neutron stars during the late inspiral phase. In the case of a nondetection of a post-merger signal following a binary neutron star inspiral we show that we can place upper limits on the energy emitted.

\end{abstract}

\pacs{}

\maketitle

\section{Introduction}
\label{sec:intro}

The coalescence of two neutron stars (NSs) emits gravitational and electromagnetic radiation (see Refs.~\cite{lrr-2012-8,Baiotti:2016qnr,2016arXiv161203050P} for reviews), providing us with a powerful probe of the NS equation of state (EoS), the properties of which are still not completely understood~\cite{Lattimer2016,Oezel2016,Oertel2017}. The first such event was recently observed~\cite{TheLIGOScientific:2017qsa,Abbott:2017eaw}. The coalescence consists of a \emph{premerger} and a \emph{post-merger} phase, both potentially observable by the ground-based gravitational wave (GW) detectors advanced LIGO (aLIGO)~\cite{0264-9381-32-7-074001} and advanced VIRGO (AdV)~\cite{TheVirgo:2014hva}. 

In the premerger phase the two NSs orbit around each other, gradually losing orbital energy and angular momentum through gravitational wave emission, speeding up, tidally deforming their companions, and eventually merging~\cite{lrr-2014-2}. The NS tidal deformation during this phase leaves an imprint on the GW emitted~\cite{2008PhRvD..77b1502F} which depends on the EoS. This imprint has been studied as a potential probe of the EoS~\cite{PhysRevLett.111.071101,2015arXiv150305405A,2014PhRvD..89j3012W,2015PhRvD..91d3002L,Chatziioannou:2015uea} suggesting that it is possible to measure the NS radius to within 1.3km for a signal emitted at 300 Mpc~\cite{2013PhRvD..88d4042R}.

After the collision the remnant evolves to a quasistable or stable state emitting additional gravitational radiation. The nature of the merger remnant depends on the component masses and on the NS EoS. Massive systems likely undergo prompt collapse to a black hole (BH) immediately after the merger. The BH remnant emits quasinormal-mode ringdown gravitational radiation which lies at frequencies $\sim 6$kHz, above the calibrated range of current and planned detectors~\cite{shibata:06bns,PhysRevD.78.084033}. For most candidate EoS a merger with typical binary masses is expected to result in a quasistable \emph{hypermassive} NS (HMNS) supported by differential rotation and thermal effects~\cite{2000ApJ...528L..29B}. The HMNS may survive for tens to hundreds of milliseconds, emitting GWs with frequencies in $(1.5-4)$kHz~\cite{1994PhRvD..50.6247Z,1996A&A...311..532R,2005PhRvL..94t1101S,2005PhRvD..71h4021S,shibata:06bns,2007PhRvL..99l1102O,2011MNRAS.418..427S,2011PhRvD..83l4008H,2012PhRvL.108a1101B,bauswein:12,2013PhRvL.111m1101B,PhysRevD.78.084033,2011PhRvL.107e1102S,hotokezaka:13,2014PhRvL.113i1104T,2014arXiv1412.3240T,2015arXiv150401764B,bauswein:15,Foucart2016,Lehner:2016lxy,Kawamura2016,2016CQGra..33x4004E,Radice:2016rys,Dietrich2017,Maione2017}, a promising bandwidth for aLIGO/AdV. For sufficiently low binary masses and depending on the exact EoS the remnant may be a supramassive NS -in which case collapse will occur after differential rotation has ceased- or a stable NS. 

Systematic studies of numerical binary NS (BNS) simulations suggest that transient nonaxisymmetric deformations and quadrupolar oscillations of the HMNS yield a short-duration high-frequency GW signal that can be used to constrain the NS EoS, e.g. Refs.~\cite{1994PhRvD..50.6247Z,1996A&A...311..532R,2005PhRvL..94t1101S,2005PhRvD..71h4021S,shibata:06bns,2007PhRvL..99l1102O,2011MNRAS.418..427S,2011PhRvD..83l4008H,2012PhRvL.108a1101B,bauswein:12,2013PhRvL.111m1101B,PhysRevD.78.084033,2011PhRvL.107e1102S,hotokezaka:13,2014PhRvL.113i1104T,2014arXiv1412.3240T,2015arXiv150401764B,bauswein:15,Foucart2016,Lehner:2016lxy,Kawamura2016,2016CQGra..33x4004E,Radice:2016rys,Dietrich2017,Maione2017} in a way that is complementary to constraints obtained from the premerger signal. In particular, it has been proposed to employ the dominant oscillation frequency to determine radii of NSs~\cite{2012PhRvL.108a1101B,bauswein:12}. Studying the post-merger phase is complementary in the sense that the post-merger phase probes a density regime of the EoS that is higher than typical densities in the merging stars. The central density of the merger remnant typically exceeds the central density of the progenitor stars. Moreover, the merger remnant may provide a way to study temperature effects of high-density matter.

\begin{figure}[h!]
\includegraphics[width=0.9\columnwidth,clip=true]{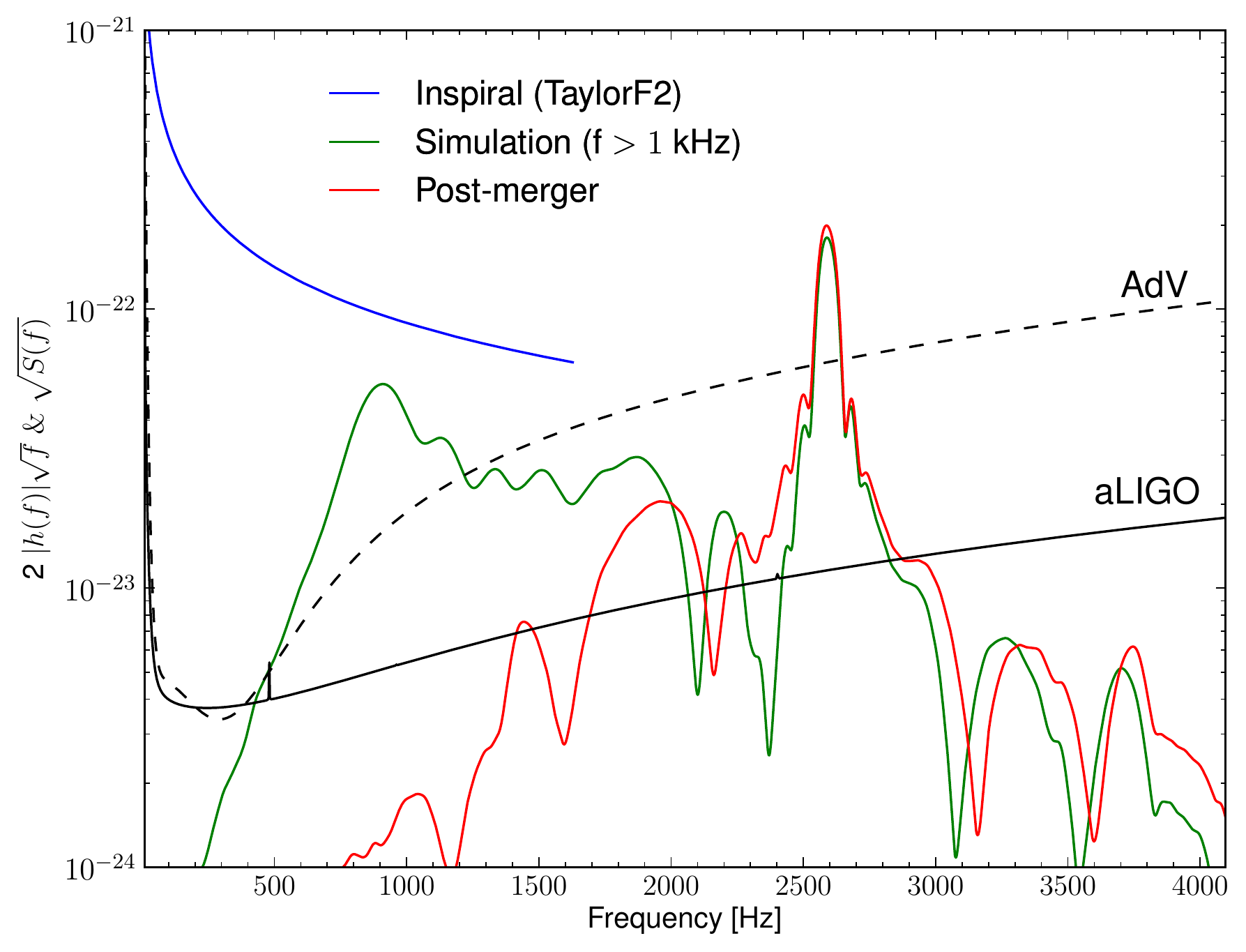}
\caption{ \label{fig:examplewaveform} Spectrum for a GW emitted during the coalescence of two nonspinning NSs with masses $1.35M_{\odot}$ at a fiducial distance of $20$Mpc, optimally oriented and with the DD2 EoS. We show the premerger point-particle phase (blue), the full simulation starting at $1$kHz (green), the post-merger phase only (red), and the expected detector sensitivity (black). Both the full simulation and the post-merger spectrum exhibit the characteristic dominant peak at about $f_{\peak}=2586$Hz.}
\end{figure}

An example spectrum for the GW emitted from a nonspinning, equal-mass BNS at the fiducial distance of $20$Mpc is shown in Fig.~\ref{fig:examplewaveform}. The binary merger was simulated with a relativistic smooth particle hydrodynamics code adopting a spatially conformally flat metric in Ref.~\cite{bauswein:14} and assuming NS matter is described by a moderate EoS, DD2~\cite{2010NuPhA.837..210H,2010PhRvC..81a5803T}. The spectrum of the full simulation data is shown in green, while the spectrum of the post-merger phase only is shown in red. Both spectra demonstrate a characteristic peak at a frequency approximately equal to the fundamental quadrupolar mode of the HMNS~\cite{2011MNRAS.418..427S,bauswein:july15}, showing that it is a true feature of the post-merger spectrum. For reference, we also plot the spectrum of the corresponding point-particle inspiral phase  (blue) and the sensitivity of the detectors (black).

The frequency of the peak of the post-merger spectrum $f_{\peak}$ has been found to correlate with quantities that characterize the NS EoS such as NS radii~\cite{2012PhRvL.108a1101B,bauswein:12}. References~\cite{2012PhRvL.108a1101B,bauswein:12} show that the peak frequency scales with the radius. For instance, for a total binary mass of $2.7M_\odot$ a particularly tight relation between $f_{\peak}$ and the radius of a $1.6M_{\odot}$ nonrotating NS ($R_{1.6}$) was found~\cite{2012PhRvL.108a1101B}. Similar relations hold for other binary masses~\cite{2012PhRvL.108a1101B,bauswein:july15}. Moreover, it is possible to relate the dominant post-merger oscillation frequency to other stellar properties of NSs, which scale in a similar manner with $f_{\peak}$ (e.g. Refs.~\cite{bauswein:12,2014arXiv1412.3240T,2015arXiv150401764B}). All these empirical relations can be used to translate a measurement of the peak frequency to a measurement of a quantity that can directly constrain the EoS.

Despite this high potential for EoS constraints, GW data analysis aspects of the post-merger signal remain less well studied compared to the premerger ones~\cite{2014PhRvD..90f2004C,Clark:2015zxa,Bose:2017jvk,Yang:2017xlf}. The post-merger phase's level of complexity would require unreasonably high computational cost to model efficiently, a prerequisite for the standard GW information extraction technique of \emph{matched filtering}. Ideally, in matched filtering one would use some physically parametrized and phase-coherent waveform model as a template. Given the absence of such a physical model one must resort to more approximate methods. An approach would be to adopt a relatively simple phenomenological model based on numerical simulations~\cite{Clark:2015zxa,Bose:2017jvk}. While such phenomenological models can offer considerable sensitivity, they are inevitably reliant on state-of-the art simulations and are incapable of identifying unmodeled or unexpected waveform phenomenology\footnote{Alternatively, one may use frequentist excess power methods such as Ref.~\cite{Klimenko:2015ypf} designed to detect signals with no knowledge of waveform morphology.  Although such maximum-likelihood methods are computationally cheap, we seek to construct the full posterior distribution on the waveform and any derived quantities, and to avoid heuristic thresholds implicit in the identification of excess power.}.

In this paper, we instead analyze the post-merger signal making only minimal assumptions on the waveform morphology. We use an existing Bayesian data analysis algorithm, {\tt BayesWave}~\cite{Cornish:2014kda,Littenberg:2014oda}, and employ its morphology-independent approach to reconstruct the post-merger GW signal through a sum of appropriate basis functions. For the basis function, we use sine-Gaussians wavelets, known as Morlet-Gabor wavelets. Both the number and the parameters of the wavelets are marginalized over using a reversible jump Markov chain Monte Carlo transdimensional sampler~\cite{doi:10.1093/biomet/82.4.711}.
 
The advantage of using {\tt BayesWave} to study the post-merger signal is threefold. First, the flexibility of the signal model allows us to reconstruct signals of generic morphology without relying on numerical simulations which sparsely cover the parameter space. Second, the use of a transdimensional sampler enables {\tt BayesWave} to marginalize over not only the parameters of the wavelets but also their number. As a consequence, {\tt BayesWave} will not overfit the data. Finally, we use a broadly tested data analysis algorithm that is a standard tool for aLIGO/AdV data analysis. This enables us to study information extraction from a post-merger signal using tools that would be applied to such detections in the future, making our results a realistic forecast.
 
We use numerical waveforms from Refs.~\cite{bauswein:12,bauswein:14,bauswein:15,bauswein:july15} to simulate GW signals and employ {\tt BayesWave} to reconstruct the observed signal, extract its peak frequency, and measure the NS radius. We find that the bounds on the NS radius obtained by the post-merger signal are competitive with their premerger counterparts with a GW detector network operating at design sensitivity. We find that statistical uncertainty leads to bounds on the NS radius of the order of 100 m for a signal emitted at 20 Mpc. If, on the other hand, we marginalize over the systematic error of the relation between the peak frequency and the radius we obtain a bound on the NS radius to within $(200-500)$m regardless of the strength of the signal, assuming {\tt BayesWave} can reconstruct it in the first place. Even though the exact projected bounds depend on the EoS and the details of the numerical simulations that impact the exact GW amplitude, we obtain radius bounds which are of the same order of magnitude as bounds derived from the premerger signal.

We stress that numerical simulations data are only employed as representative signals and are not used to specifically tune the reconstruction algorithm. The rest of the paper describes the details of our analysis. In this work the total binary mass refers to the sum of the gravitational mass of the binary components at infinite orbital separation.

\section{Analysis Method}
\label{sec:analysismethod}

The objective of GW inference is to determine the properties of an incident signal. In the Bayesian framework we calculate $p(h|d)$, the posterior distribution function for the signal $h$ in data $d$. Bayes' theorem links the posterior for the signal to a prior distribution function $p(h)$ and a likelihood function $p(d|h)$ through
\begin{equation}
p(h|d)=\frac{p(h)p(d|h)}{p(d)},
\end{equation}
where $p(d)$ is the evidence, and the likelihood encodes all new information we obtain from the data. The standard assumption of stationary and Gaussian data leads to a well-studied and generally accepted form for the likelihood function~\cite{Veitch:2014wba}. The prior for the signal $p(h)$ quantifies our assumptions for the GW signal.  

When studying GW signals for which accurate models exist, the signal prior demands that the GW matches the waveform model exactly; $p(h)=\delta[h-h(\vec{\theta})]p(\vec{\theta})$, where $h(\vec{\theta})$ is some parametrized GW model, and $\vec{\theta}$ are its parameters. Examples of such models are the phenomenological inspiral-merger-ringdown models~\cite{Hannam:2013oca} or the effective-one-body models~\cite{Bohe:2016gbl} used for the analysis of binary BH systems. These models are parametrized in terms of the physical parameters of the underlying system, such as the masses and the spins of the coalescing bodies. These parametrizations encompass very restrictive prior assumptions, and hence deliver the most precise results but are only accurate in the restricted regime where the assumptions about the source are reasonable.

When the GW signal is not understood well enough a more flexible parametrization for the signal is needed. One such prior can be obtained by expressing the signal as a sum of functions $w_i(\vec{y})$ with parameters $\vec{y}$; 
\begin{align}
p(h)=\delta \left[h-\sum^N_i w_i(\vec{y})\right]p(N,\vec{y}).\label{eq:priors}
\end{align}
Despite demanding that the signal matches the model exactly, this prior can be rendered very flexible depending on the choice of basis functions. If, for example, we select $N=1$ and $w(\vec{y})$ to be a binary BH template we recover the template-based analysis previously described. If, on the other hand, $N$ is allowed to vary and the $w_i(\vec{y})$ are chosen from some appropriate basis, the signal model is flexible enough to describe signals of arbitrary morphology.

The choice of basis functions is instrumental in constructing an analysis that is both flexible and efficient. For this study we work with {\tt BayesWave}, a Bayesian algorithm that decomposes the GW signal in Morlet-Gabor wavelets~\cite{Cornish:2014kda,Littenberg:2014oda}, achieving robust identification and reconstruction of morphologically uncertain GW signals~\cite{Littenberg:2015kpb,2016PhRvD..93b2002K,Becsy:2016ofp,PhysRevD.93.122004}. GW signals are modeled at the geocenter as an elliptically polarized superposition of an arbitrary number of Morlet-Gabor wavelets
\begin{eqnarray}
h_+(t) & = & \sum_{i=0}^{N_{\mathrm{s}}} \Psi(t; A_i, f_{0,i}, Q_i, t_{0,i},\phi_i)\nonumber \\
h_{\times}(t) & = & \epsilon h_+(t) e^{i\pi/2},
\end{eqnarray}
where $\epsilon$ is the ellipticity parameter, $Q\equiv 2\pi f_0 \tau$. Each wavelet depends on five parameters: an overall amplitude $A$, a quality factor $Q$, a central frequency $f_0$, a central time $t_0$, and a phase offset $\phi_0$; 
\begin{equation}
\Psi(t; A, f_{0}, \tau, t_{0},\phi_0) = A e^{-(t-t_0)^2/\tau^2} \cos{[2\pi f_0(t\!-\!t_0)\!+\!\phi_0]}.
\end{equation}
The frequency-domain strain induced in a given detector is
\begin{equation}
h(f) = \left[F^+(\theta, \phi, \psi) h_+(f) + F^{\times} (\theta, \phi, \psi) h_{\times}(f)\right]e^{2\pi i \Delta t(\theta, \phi)},
\end{equation}
where $F_{+},~F_{\times}$ are detector antenna patterns given a sky location $(\theta,\phi)$ and polarization angle $\psi$, and $\Delta t(\theta, \phi)$ is a sky location-dependent time shift relative to the time of arrival at the geocenter.

{\tt BayesWave} employs a reversible jump Markov chain Monte Carlo algorithm to sample the joint posterior of the sky location, polarization angle, ellipticity, and number $N_{\mathrm{s}}$ and parameters $(A_i, f_{0,i}, Q_i, t_{0,i},\phi_i)$ of the wavelets. The samples are then used to produce draws from the waveform posterior $p(h|d)$ itself. Subsequently using the waveform samples one can derive posteriors on quantities that describe features of the waveform such as the frequency of the peak of the spectrum.

The use of a transdimensional sampler to determine the number of wavelets in the reconstructed signal ensures that {\tt BayesWave} does not overfit the data. In practice, adding a wavelet to the signal reconstruction increases the dimensionality of the model, incurring an Occam-type reduction in the posterior probability. As a result, the additional wavelet will only be retained in the reconstruction if it improves the fit to the data considerably so as to overcome the Occam penalty.

As can be seen from Eq.~\eqref{eq:priors} the priors of the analysis refer to the number and parameters of the individual wavelets. We study $250$ms of data in the $(1024,4096)$Hz frequency range. This range was chosen such that it includes most of the post-merger emission from both soft and stiff EoS. A consequence of this frequency range is that most of the signals we are analyzing include both the merger and post-merger phases; see Fig.~\ref{fig:examplewaveform}. For this reason we impose a minimum number of two wavelets used, while the prior on the quality factor $Q$ is flat between $1$ and $200$. We employ the prior proposed and discussed in Ref.~\cite{Cornish:2014kda} for the wavelet amplitude. Finally, the prior on the wavelet phase offset is uniform between $0$ and $2\pi$.

The quality of the reconstruction is described through the \emph{overlap} between signal $s$ and model $h$;
\begin{equation}
{\cal{O}} \equiv \frac{\langle s,h \rangle}{\sqrt{\langle s,s \rangle} \sqrt{\langle h,h \rangle}},
\end{equation}
while the strength of the signal is quantified through the signal-to-noise ratio (SNR);
\begin{equation}
\text{SNR} \equiv \langle s,s \rangle.
\end{equation}
In the above equations we have defined the inner product
\begin{equation}
\langle a,b \rangle \equiv 4 \Re \int_{f_{min}}^{f_{max}} \frac{a(f)b^*(f)}{S_n(f)} df,
\end{equation}
where $S_n(f)$ is the detectors noise spectral density and $(f_{min},f_{max})=(1024,4096)$Hz is the bandwidth of the analysis. 

\begin{figure}[h!]
\includegraphics{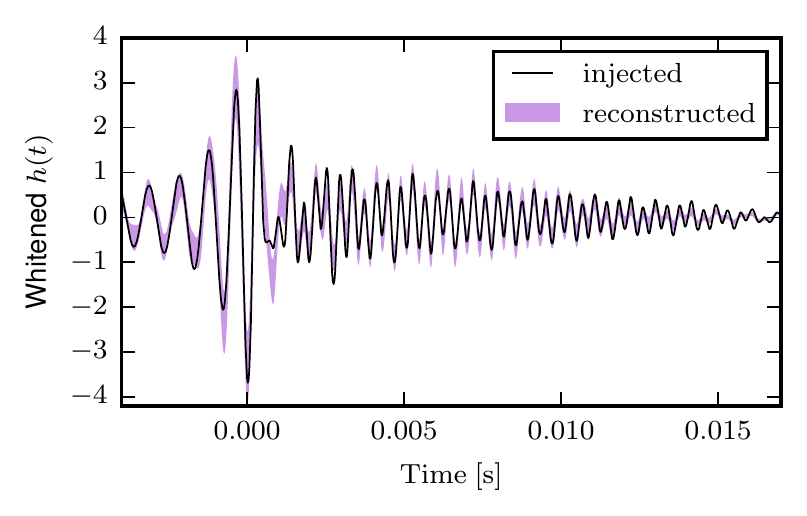}\\
\includegraphics[width=\columnwidth,clip=true]{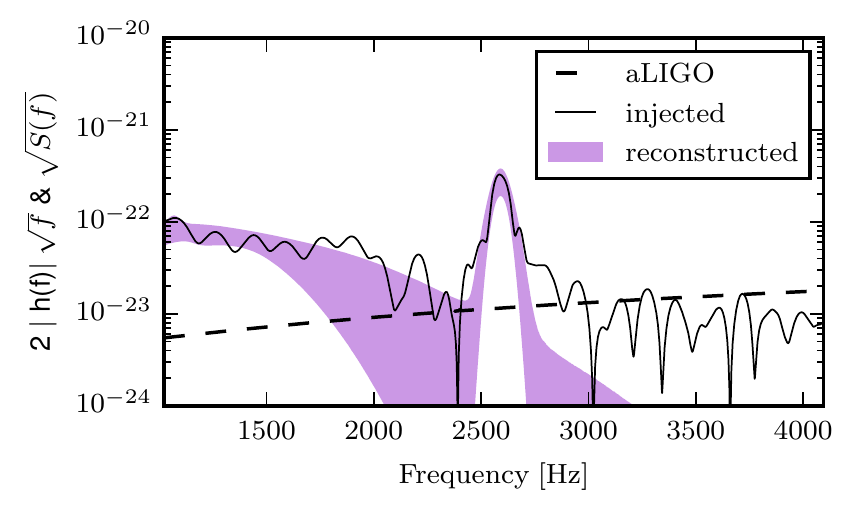}
\caption{ \label{fig:dd2_135135_reconstruction} Injected and reconstructed whitened time-domain data (top) and spectrum (bottom) for a signal produced by two nonspinning, (1.35, 1.35)M$_{\odot}$ NSs with the DD2 EoS~\cite{2010NuPhA.837..210H,2010PhRvC..81a5803T} at a post-merger SNR of $5$ -corresponding roughly to a distance of 20Mpc for an optimally oriented source- as observed by the Hanford detector. The shaded region denotes the $90\%$ CI of the reconstruction. The dashed line in the bottom panel is the detector sensitivity. The {\tt Bayeswave} reconstruction is able to capture the main features of the signal including the post-merger spectrum peak. }
\end{figure}

For a demonstration of the {\tt Bayeswave} analysis we consider the post-merger GW emission of an equal-mass BNS coalescence simulated in Ref.~\cite{bauswein:14}. Each binary component has a mass of $1.35M_{\odot}$ and the DD2 EoS~\cite{2010NuPhA.837..210H,2010PhRvC..81a5803T} was employed in the simulation. The signal is scaled to a post-merger\footnote{We define ``post-merger'' as all times after the time of peak amplitude, and the post-merger SNR is computed by truncating and windowing the waveform in the time domain.} SNR of 5, assuming the design sensitivity of aLIGO~\cite{AdvLIGO-noise} and AdV~\cite{TheVirgo:2014hva}. The short duration $\sim 10$ms of the GW signal makes it ideal for model-agnostic algorithms of which the performance deteriorates as the time-frequency volume of the search space increases\footnote{In principle, the longer duration signals emitted from remnants that survive for hundreds of milliseconds before collapse could also be analysed with {\tt Bayeswave} depending on their time-frequency signature. We plan to explore more types of signals in the future.}.

\begin{figure}[h!]
\includegraphics[width=\columnwidth,clip=true]{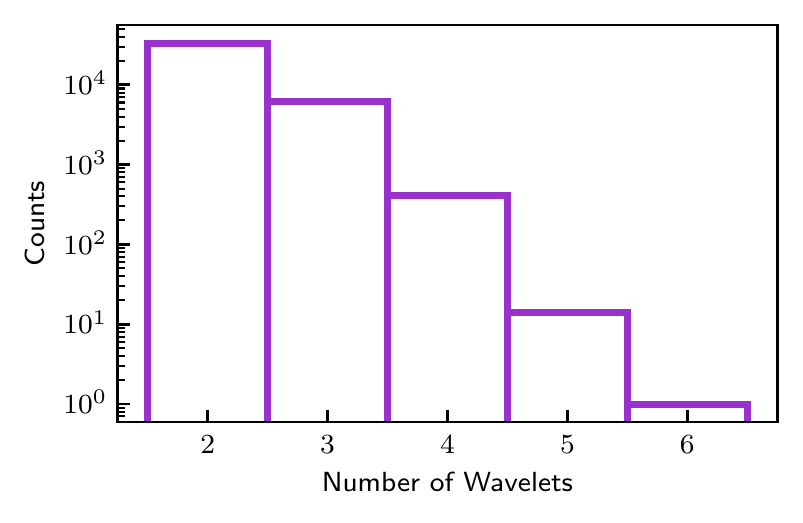}
\caption{ \label{fig:model_dimensions} Histogram of the number of wavelets  {\tt BayesWave} used for the signal reconstructions of Fig.~\ref{fig:dd2_135135_reconstruction}. {\tt BayesWave} uses model selection to determine the most probable number of wavelets.}
\end{figure}

We use this numerical waveform to simulate data~\cite{Schmidt:2017btt} and inject it in a network of two aLIGO detectors and AdV at design sensitivity and reconstruct the signal with {\tt BayesWave}. Figure~\ref{fig:dd2_135135_reconstruction} shows the posteriors for the whitened time-domain (top panel) and spectrum (bottom panel) reconstructions. Both plots show the injected signal (black), and the 90\% credible interval (CI) of the reconstruction posterior (magenta). Figure~\ref{fig:model_dimensions} shows a histogram of the number of wavelets used for this reconstruction; {\tt BayesWave} used $\sim(2-3)$ wavelets to achieve the reconstruction of Fig.~\ref{fig:dd2_135135_reconstruction}.

These plots demonstrate how {\tt BayesWave} is capable of reconstructing the dominant features of the injected signal, including the dominant post-merger frequency with only \emph{minimal} assumptions about the signal morphology. On the other hand, the absence of a matched filter means that {\tt BayesWave} does not reconstruct the entire signal, but only its most prominent features. We study the reconstruction performance and its relation to the strength of the signal in the following section.

\section{Reconstruction Performance}
\label{sec:reconstruction}

In this section we systematically study the reconstruction performance of {\tt BayesWave} for signals of different strengths and EoS. We select three representative EoS (NL3~\cite{1997PhRvC..55..540L,2010NuPhA.837..210H} for stiff, DD2~\cite{2010NuPhA.837..210H,2010PhRvC..81a5803T} for moderate, and SFHO~\cite{2013ApJ...774...17S} for soft) and use numerical waveforms from Refs.~\cite{bauswein:14,bauswein:july15} to simulate signals in a network of two aLIGOs and AdV at design sensitivity\footnote{Alternative networks with better sensitivity such us tuned configurations~\cite{AdvLIGO-noise}, squeezing~\cite{PhysRevD.91.044032}, or 3rd generation detectors~\cite{Hild:2010id,0264-9381-34-4-044001} would yield better results than the ones presented here and are left for future work.}. All simulated signals in this section have the same intrinsic and extrinsic parameters but the EoS and the distance/SNR. The system parameters were chosen such that they lead to results similar to a typical binary system, as demonstrated by the Monte Carlo analysis of Sec.~\ref{sec:montecarlo}. We consider the results of this section as ``representative'' of a larger population. The injections do not contain a specific noise realization, as this has been shown to be equivalent to averaging over noise realizations~\cite{Nissanke:2009kt}.

For reference, a BNS with the moderate DD2 EoS at 20Mpc in a network of two aLIGOs and AdV at design sensitivity has a maximum post-merger SNR of about 5 and an orientation-averaged SNR of about 1. These SNR values are higher (lower) for stiff (soft) EoS. Recall that the SNR scales inversely with the distance and the current aLIGO/VIRGO sensitivity is expected to be a factor of a few below the design one~\cite{Aasi:2013wya}. More detailed calculations for the correspondence between distance and SNR are presented in Table II of Ref.~\cite{Clark:2015zxa}.

\subsection{Overlap}
\label{sec:overlap}

\begin{figure}
\includegraphics[width=0.9\columnwidth,clip=true]{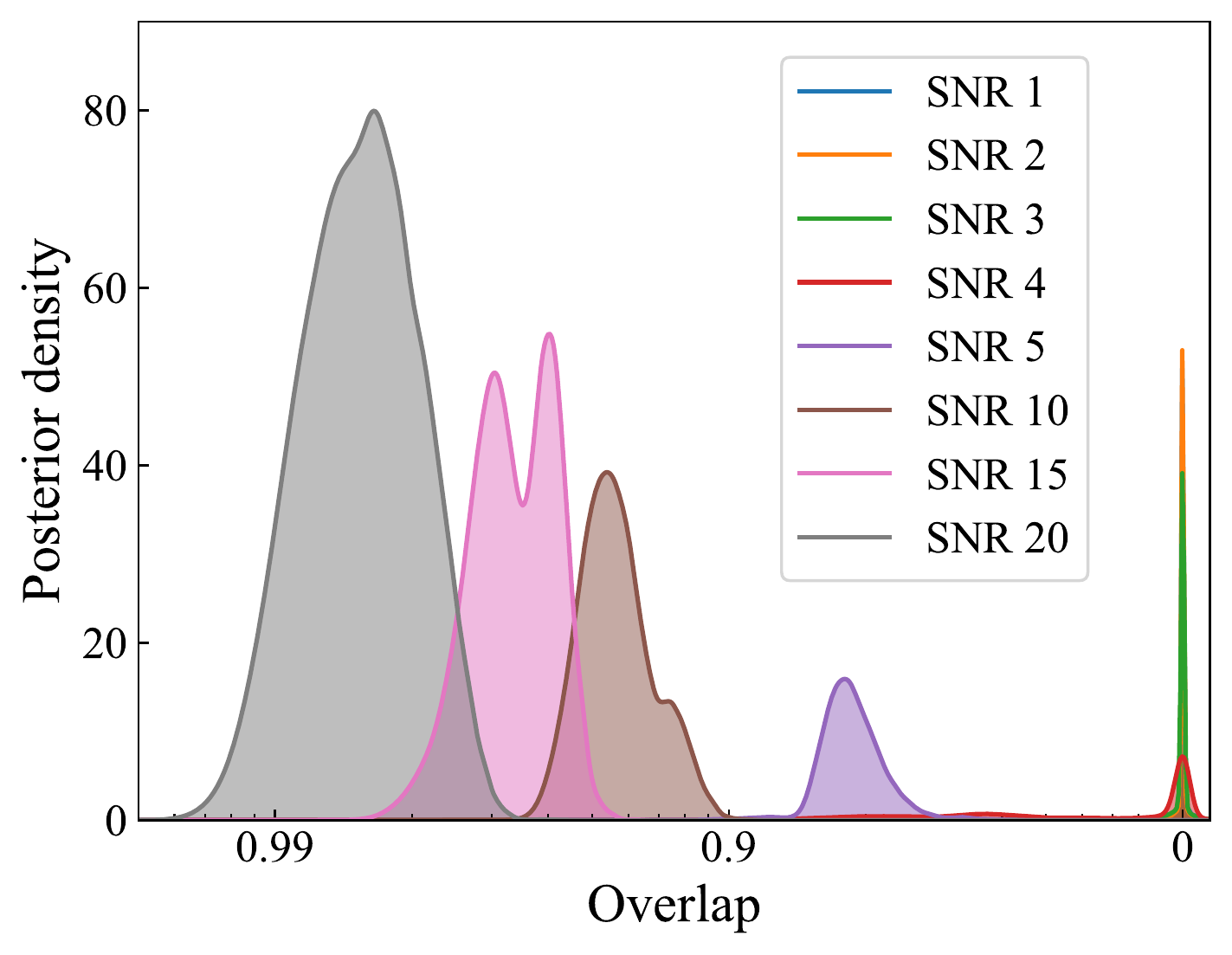}\\
\includegraphics[width=0.9\columnwidth,clip=true]{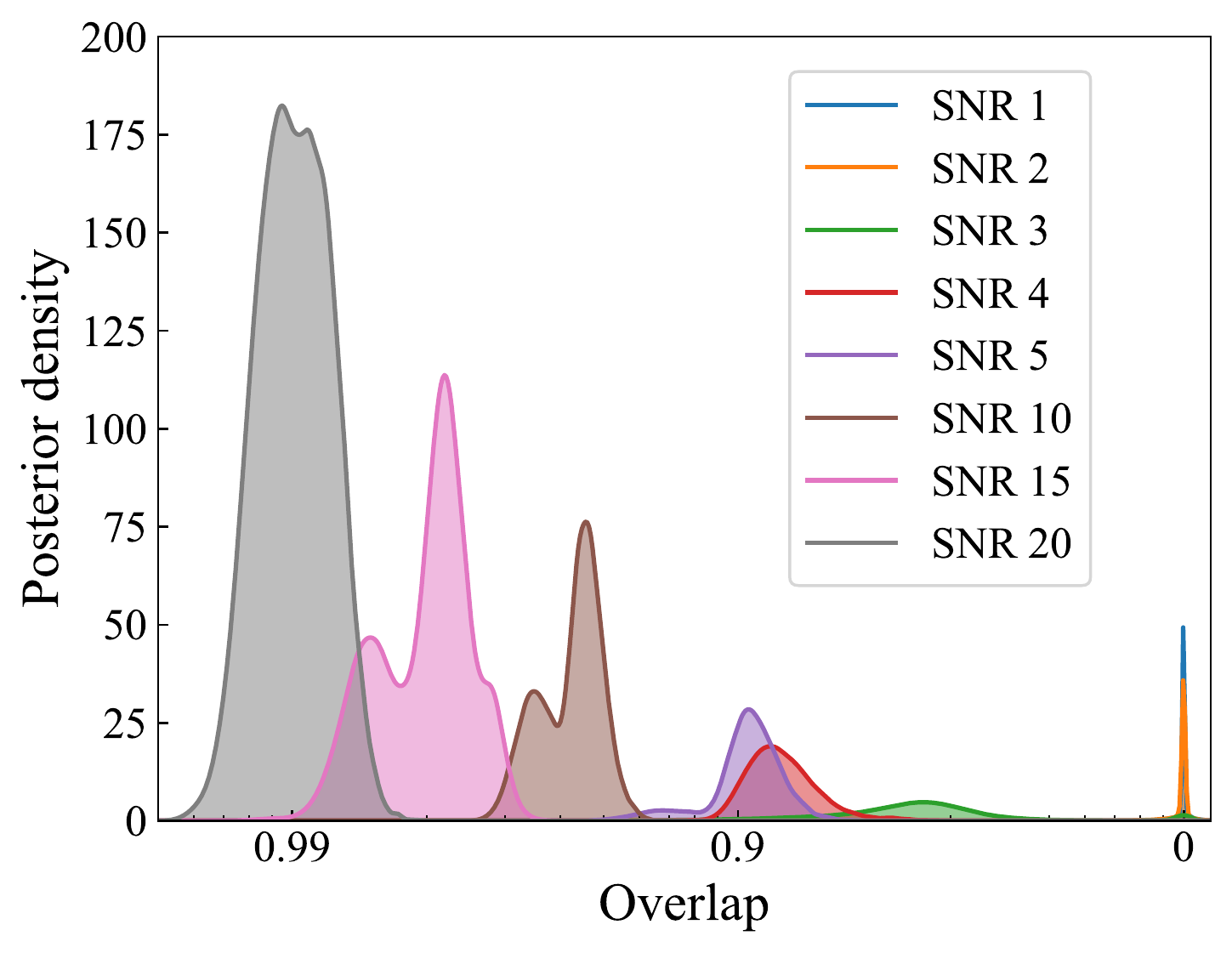}\\
\includegraphics[width=0.9\columnwidth,clip=true]{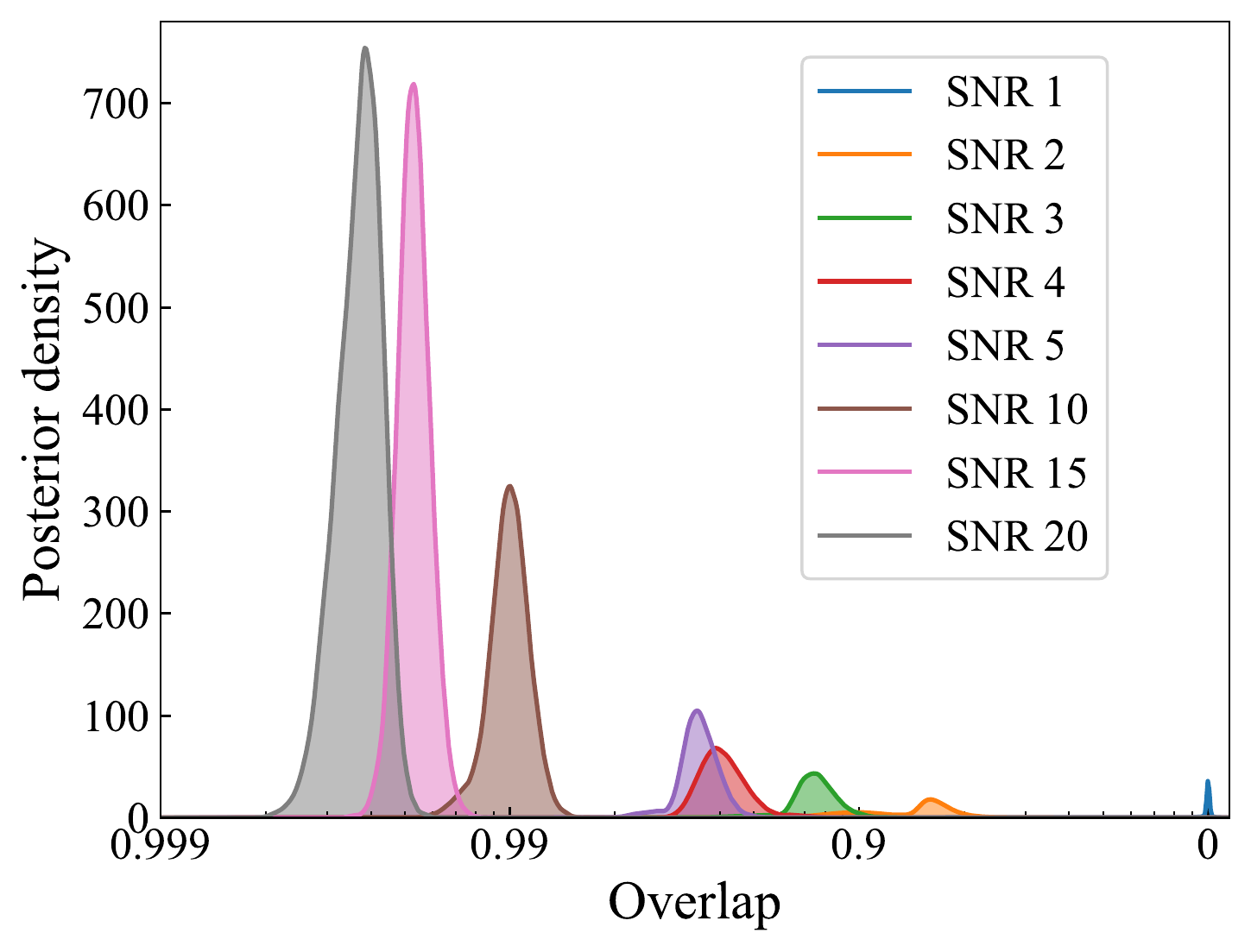}
\caption{\label{fig:overlapposteriors} Overlap posterior density function for NL3 (top), DD2 (middle), and SFHO (bottom) for different post-merger SNR values. As the SNR of the injected signal increases, {\tt BayesWave} achieves more faithful reconstructions of the signal.}
\end{figure}

The injected signals are analyzed with {\tt BayesWave} and Fig.~\ref{fig:overlapposteriors} shows the posterior distribution of the overlap between the injected and the reconstructed signal for each EoS. Recall that the overlap quantifies how faithful the reconstructed signal is to the true injected one, with an overlap of $1$ denoting perfect reconstruction.

As the post-merger SNR of the injected signal increases the overlaps {\tt BayesWave} achieves an increase too, signaling more accurate reconstructions. This is a demonstration of the inherent trade-off between goodness of fit and simplicity of Bayesian inference. In order for  {\tt BayesWave} to improve the overlap and the reconstruction, it needs to use more wavelets. Since the addition of each wavelet increases the dimensionality of the model, the resulting Occam penalty can only be overcome if the wavelet helps improve the fit considerably. On the other hand, if the extra wavelet does not help improve the fit enough, the reconstruction will be disfavored. This process shields {\tt BayesWave} from overfitting the data.

The overlap does not reach its nominal maximum value of $1$ (perfect reconstruction), which means that {\tt BayesWave} does not fully reconstruct the injected signal. However, the overlap values achieved are above $90\%$ for post-merger SNRs above $\sim5$, making this analysis at least competitive with existing phenomenological models~\cite{Bose:2017jvk,Yang:2017xlf} without suffering from systematic uncertainties from over-relying on uncertain numerical simulations.

\subsection{Peak Frequency}
\label{sec:fpeak}
\begin{figure}[h!]
\includegraphics[width=0.9\columnwidth,clip=true]{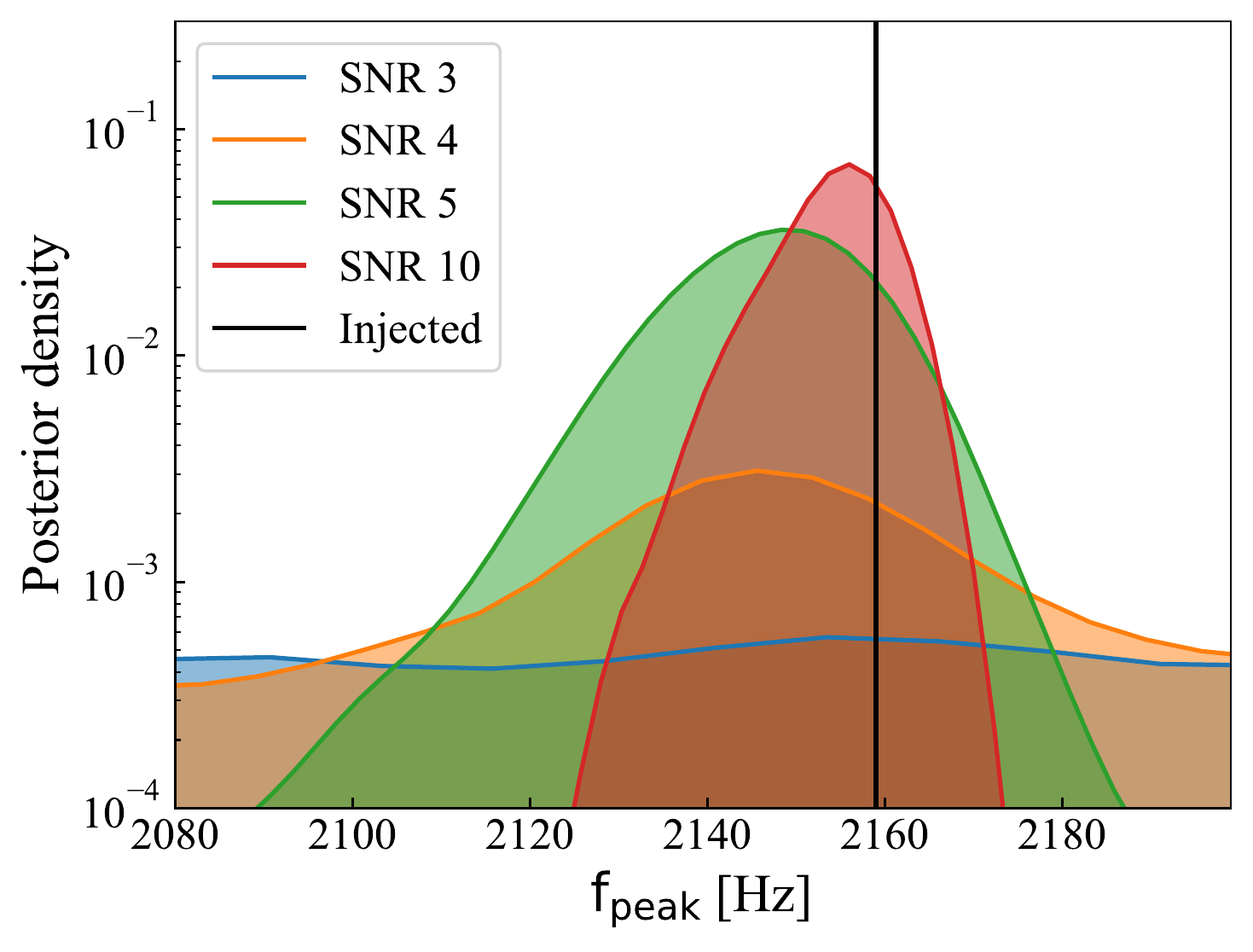}\\
\includegraphics[width=0.9\columnwidth,clip=true]{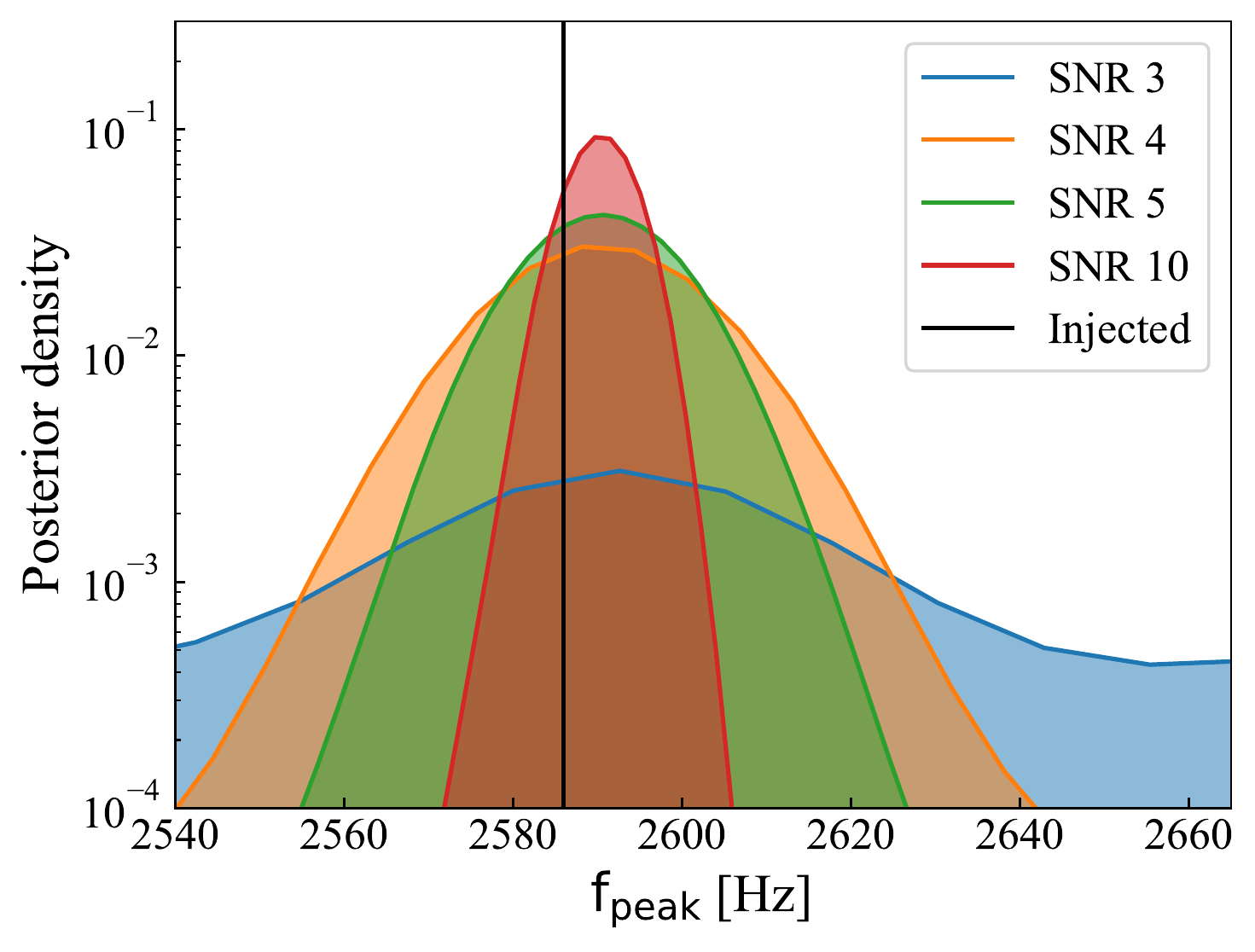}\\
\includegraphics[width=0.9\columnwidth,clip=true]{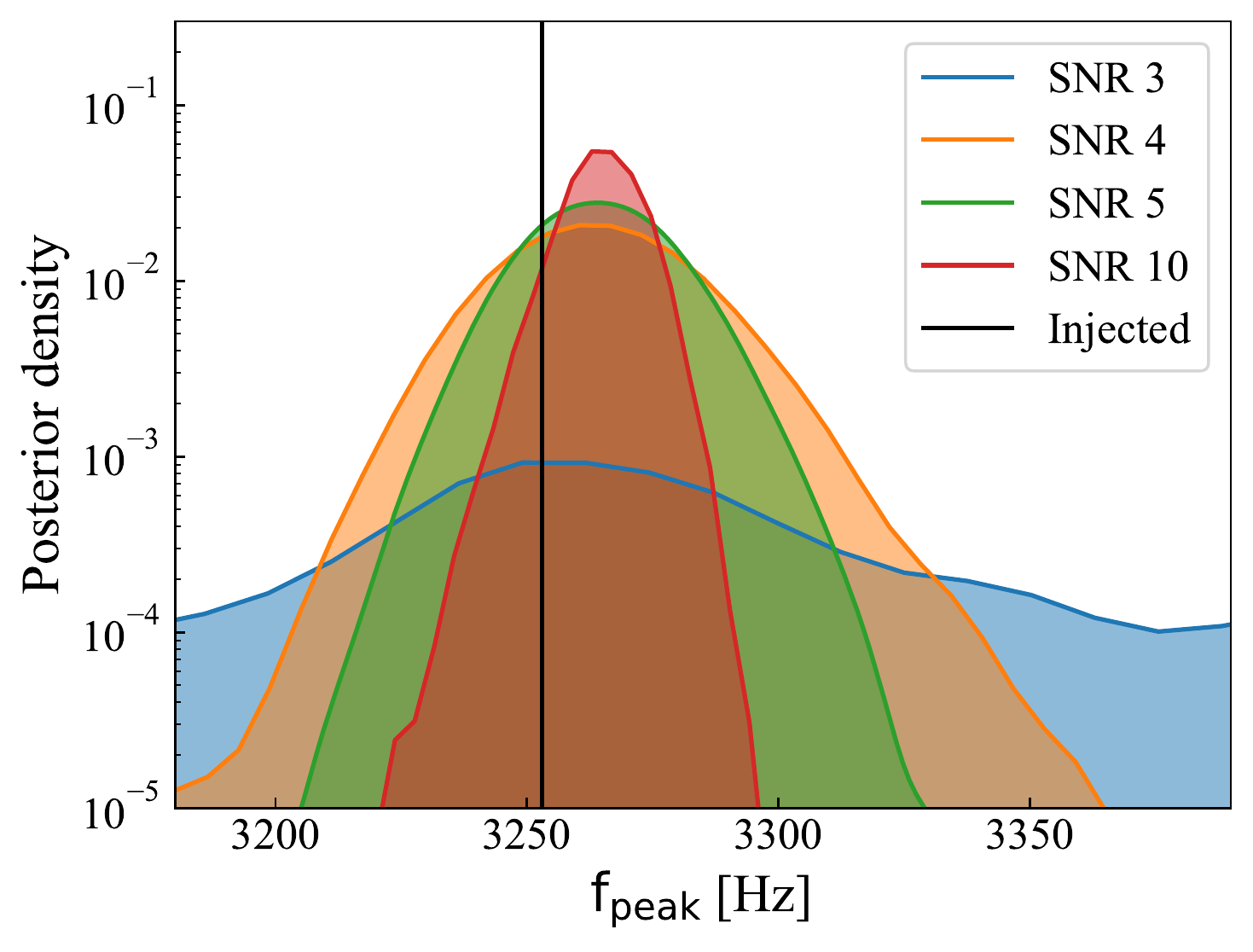}
\caption{\label{fig:fpeakposteriors}Peak frequency posterior density function for NL3 (top), DD2 (middle), and SFHO (bottom) for different post-merger SNR values. The vertical line denotes the correct (injected) value. At low SNR the posterior for the peak frequency is uninformative and similar to the prior. As the SNR increases, {\tt BayesWave} achieves a more accurate reconstruction of the signal and the posterior peaks at the correct value for $f_{\peak}$.}
\end{figure}

The posterior for the reconstructed signal (see for example Fig.~\ref{fig:dd2_135135_reconstruction}) can be used to calculate the posterior for the dominant post-merger frequency $f_{\peak}$. For each sample in the posterior for the reconstructed signal we suppress the inspiral and merger phases by applying a window at the measured maximum time-domain amplitude.  We then define $f_{\peak}$ as the frequency of the maximum of the post-merger spectrum in the range $[1500,4000]$Hz\footnote{Despite expecting post-merger power as low as $\sim1$kHz, the peak frequency is expected in the $(1500-4000)$Hz range.}. If a certain reconstruction sample does not possess a maximum, then instead we draw a sample from $f_{\peak}$'s prior distribution function. Overall the posterior distribution function for $f_{\peak}$ is
\begin{equation}
p(f_{\peak}|d)=(n-1)p(f_{\peak}) + n \,s(f_{\peak}|d)\label{posteriordefinition},
\end{equation}
where $n$ is the relative number of samples that possessed a peak, $p(f_{\peak})$ is the prior, and $s(f_{\peak}|d)$ is the distribution of the $f_{\peak}$ samples calculated from the reconstructed spectrum.

Figure~\ref{fig:fpeakposteriors} shows the posterior for $f_{\peak}$ for different EoS and signal strengths. At low SNR values the posterior is equal to the prior, i.e. most reconstructed spectra do not exhibit a peak. As the SNR increases the data become more informative and the posterior peaks around the correct $f_{\peak}$ value. From this plot we conclude that we can measure $f_{\peak}$ to within about $36(27)[45]$Hz at the $90\%$ credible level for a stiff(moderate)[soft] EoS at a post-merger SNR of 5.

Comparing the posterior distributions for the peak frequency to the true injected value for $f_{\peak}$ (vertical black line) reveals that there is a systematic shift between the two even for the relatively high SNR of 10. The reason for this has to do with the exact shape of the peak of the spectrum. Figure~\ref{fig:dd2_135135_reconstruction} shows that the dominant post-merger frequency in not strictly constant in time. As a result, the peak of the spectrum is not symmetric about its maximum, something that is visible in the bottom panel of Fig.~\ref{fig:dd2_135135_reconstruction}. As a consequence, the shape of the peak does not exactly match the basis function used by {\tt BayesWave}; the frequency-domain representation of Morlet-Gabor wavelets is symmetric about its maximum. This mismatch results in {\tt BayesWave} shifting the wavelet that reconstructs the spectrum peak in frequency in an effort to maximize the recovered signal, resulting in the bias seen in Fig.~\ref{fig:fpeakposteriors}.

The time evolution of the peak frequency suggests that  the constant-frequency Morlet-Gabor wavelets might not be the ideal basis function for post-merger signals. As an alternative, we studied the ``chirplets'' of Ref.~\cite{chirplets}, which are Morlet-Gabor wavelets of which the frequency is allowed to vary. This variation is encoded in an extra parameter that gives the constant time derivative of the frequency. The additional parameter increases the dimensionality of the model making it harder for {\tt BayesWave} to use many chirplets. Indeed we find that chirplets tend to reconstruct the signal less well than Morlet-Gabor wavelets; the extra parameter per chirplet forces the code to use fewer chirplets than wavelets, resulting in poorer reconstructions. We leave further exploration of other basis functions for future work.

Comparing the posteriors in the three panels of Fig.~\ref{fig:fpeakposteriors} shows that the softer the EoS, the easier it is to measure the peak frequency for signals of a constant SNR. This is because soft EoS have larger values of $f_{\peak}$ and hence accumulate more radians of the GW phase at the same amount of time, making it easier to measure the frequency. Indeed, the $f_{\peak}$ posterior becomes marginally informative at SNR $3 (3) [4]$ for the soft (moderate) [stiff] EoS.
 
\subsection{Radius}
\label{sec:radius}
\begin{figure}[h!]
\includegraphics[width=\columnwidth,clip=true]{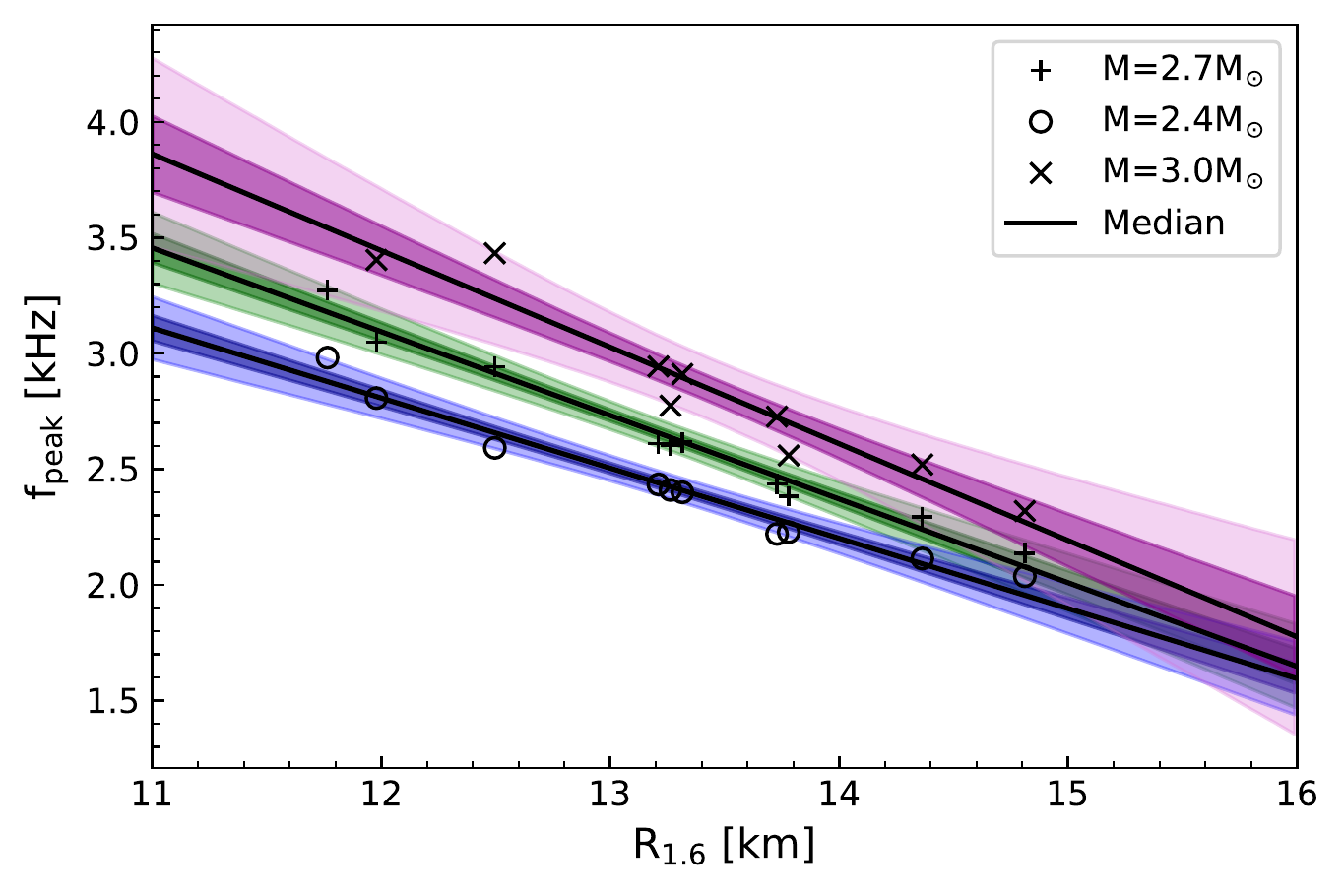}
\caption{\label{fig:fpeak_R16-intervalsplot} EoS-independent relation between the frequency of the post-merger spectrum peak and the radius of a $1.6M_{\odot}$ nonrotating NS for different total masses. The symbols are data calculated from numerical simulations of merging NSs with different total masses; each color represents the fit for data of the same total mass, the black line is the median of each fit, and the shaded regions denote the $50\%$ (dark colored) and $90\%$ (light colored) CIs of the fit.}
\end{figure}
\begin{figure}[h!]
\includegraphics[width=0.9\columnwidth,clip=true]{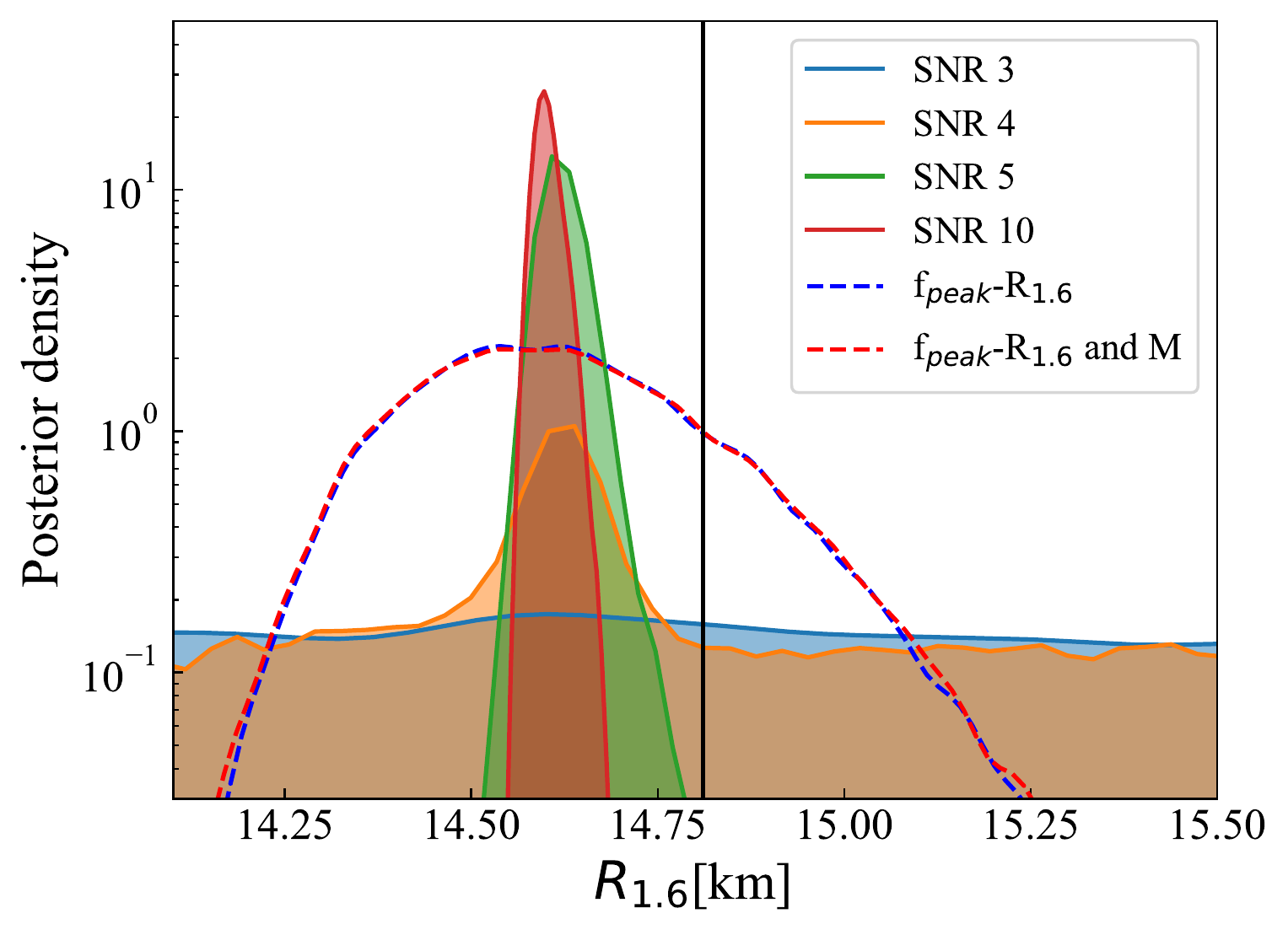}\\
\includegraphics[width=0.9\columnwidth,clip=true]{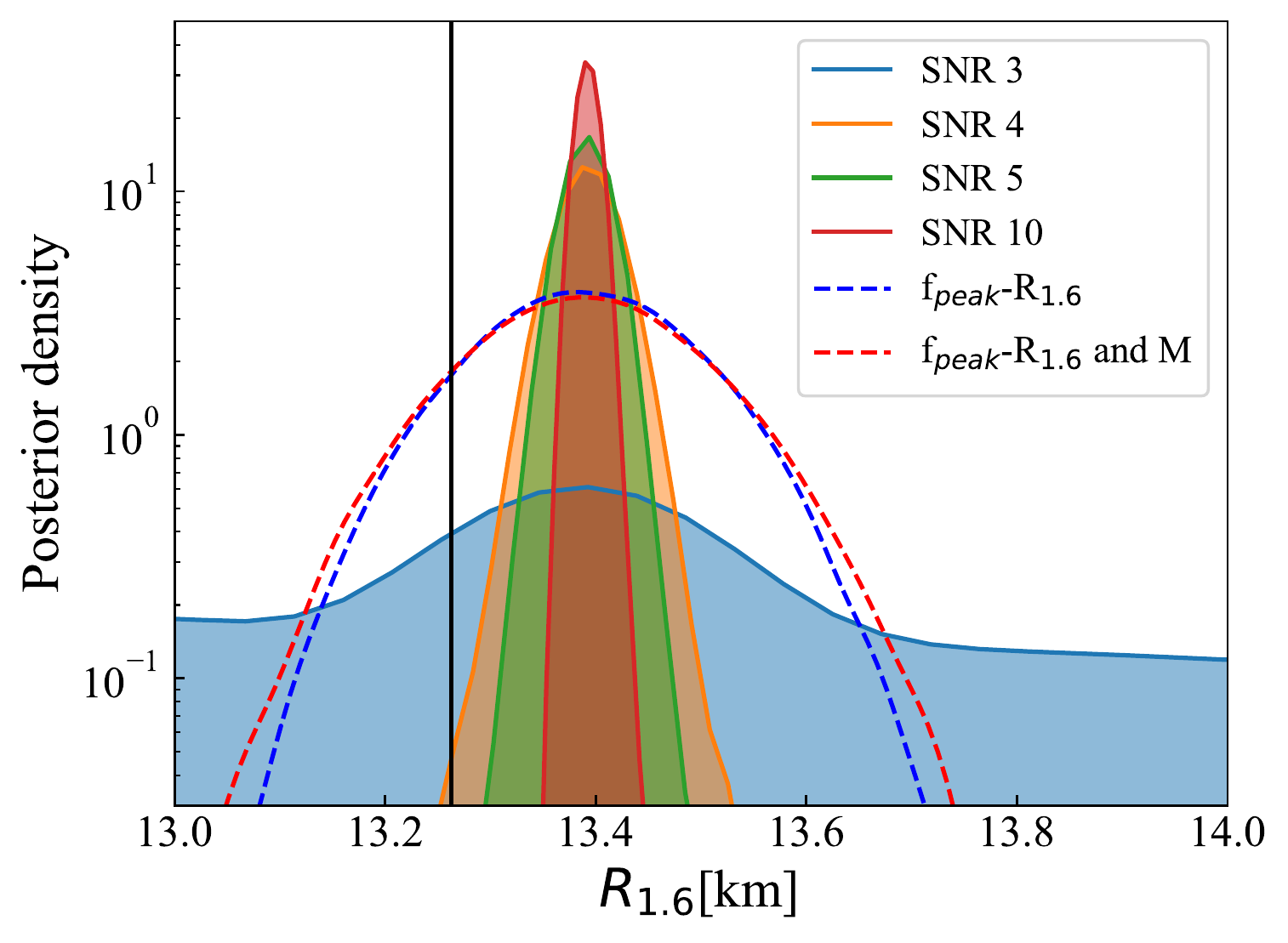}\\
\includegraphics[width=0.9\columnwidth,clip=true]{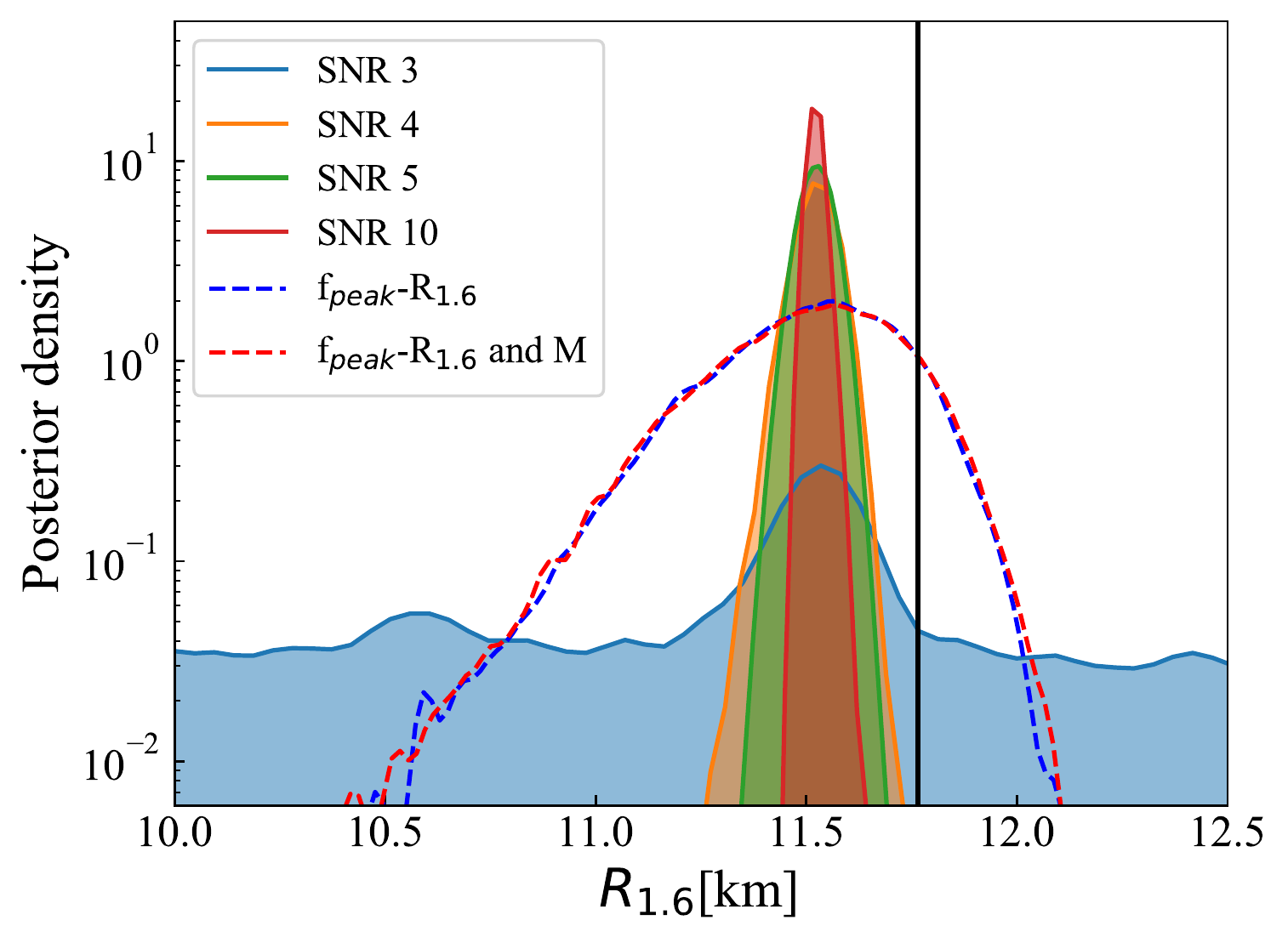}
\caption{\label{fig:radiusposteriors} Radius posterior density function for NL3 (top), DD2 (middle), and SFHO (bottom) for different post-merger SNR values. The shaded posteriors are calculated with the maximum likelihood fit to the $f_{\peak}/M-R_{1.6}$ relation shown in Fig.~\ref{fig:fpeak_R16-intervalsplot}. The nonfilled dashed posteriors are obtained by marginalizing over the uncertainty in the $f_{\peak}/M-R_{1.6}$ relation including the total mass uncertainty (red dashed) and fixing the total mass to its injected value (blue dashed).}
\end{figure}

Numerical simulations of NS coalescences have suggested that a measurement of the peak frequency can be used to constrain the NS EoS. Specifically, Ref.~\cite{bauswein:12} showed that the peak frequency of $(1.35-1.35)M_\odot$ mergers is correlated with the radius of a $1.6M_{\odot}$ nonrotating NS ($R_{1.6}$) in a way that does not depend on the underlying EoS. Therefore, a potential measurement of $f_{\peak}$ from the post-merger signal can be used to obtain an estimate on $R_{1.6}$, a quantity that can be used to directly constrain the EoS. Choosing $R_{1.6}$ to characterize the post-merger GW emission and the underlying EoS, respectively, is guided by the empirical finding that for this binary mass the frequency-radius relation shows a relatively small scatter. Other choices are possible, e.g. $R_{1.35}$ or $R_{1.8}$, yielding similar empirical relations with a potentially larger scatter. Moreover, similar relations are found for other binary masses~\cite{bauswein:12,bauswein:july15}, see Fig.~\ref{fig:fpeak_R16-intervalsplot}.

Empirical relations are not exact but exhibit an intrinsic scatter. If the deviation from exact universality is not taken into account an additional systematic error enters the analysis. In the Bayesian framework such systematic uncertainties are dealt with by modeling and marginalization. An example of this procedure is presented in Figs.~\ref{fig:fpeak_R16-intervalsplot} and~\ref{fig:radiusposteriors}, in which we convert our posteriors for the peak frequency to posteriors for $R_{1.6}$. 

Figure~\ref{fig:fpeak_R16-intervalsplot} describes the relation between $f_{\peak}$ and $R_{1.6}$ for different values of the total mass\footnote{We restrict this analysis to equal-mass binaries and leave the exploration of the exact impact of the mass ratio to future work. Adopting equal-mass systems may be an acceptable approximation if the measurement of the inspiral phase can verify a sufficiently symmetric binary configuration.}.  Symbols in the plot denote the results from numerical simulations of BNS coalescences of different total masses~\cite{bauswein:12,bauswein:14,bauswein:july15}. We divide each set of data by the total mass and fit them with a linear model and plot the best-fit and median models as well as $50\%$ and $90\%$ CIs. The extent of these intervals quantifies the deviation from universality in the $f_{\peak}/M-R_{1.6}$ relation.

With this result in hand we can estimate the posterior distribution function for $R_{1.6}$, given in Fig.~\ref{fig:radiusposteriors}. The shaded posteriors are calculated using the best-fit model from Fig.~\ref{fig:fpeak_R16-intervalsplot} as well as perfect knowledge of the total mass to convert the posteriors for the peak frequency of Fig.~\ref{fig:fpeakposteriors} into posteriors for $R_{1.6}$. This method ignores any systematic uncertainties in the $f_{\peak}/M-R_{1.6}$ relation. As expected, a precise measurement of $f_{\peak}$ leads to a tight measurement of $R_{1.6}$ to within $100(75)[120]$m at the $90\%$ credible level for a stiff(moderate)[soft] EoS at a post-merger SNR of 5.

Despite its high precision, such an $R_{1.6}$ measurement is not accurate. Ignoring the spread in the $f_{\peak}/M-R_{1.6}$ relation has resulted in a large systematic error that surpasses the statistical measurement uncertainty. The result is that the posterior measurement does not agree with the injected true value inducing a large measurement bias. If we instead marginalize over the uncertainty in the $f_{\peak}/M-R_{1.6}$ we obtain more broad posteriors that do include the injected value of $R_{1.6}$. The marginalized posteriors are included in Fig.~\ref{fig:fpeak_R16-intervalsplot} and are similar irrespective of the SNR of the signal; in red dashed we show the resulting posterior from marginalizing over the uncertainty of the $f_{\peak}/M-R_{1.6}$ relation while keeping the total mass fixed to its injected value; in blue dashed we show the resulting posterior from marginalizing both over the $f_{\peak}/M-R_{1.6}$ relation uncertainty and the total mass measurement uncertainty. For projections of the total mass measurement uncertainty we use the estimates derived in Ref.~\cite{Farr:2015lna}. We find that the total mass is determined extremely accurately from the inspiral phase (measurement error of the order of $10^{-2}-10^{-3}$) and has little effect on the resulting posterior. Despite the fact that the marginalized posteriors are significantly broader than the ones derived with the best-fit $f_{\peak}/M-R_{1.6}$ relation, we still arrive at a measurement of $R_{1.6}$ of the order of $(300-700)$m, independently of the SNR as long as {\tt BayesWave} can detect the signal.

In order to compare this measurement accuracy to constraints on the NS radius obtained from the premerger phase, we need to estimate the premerger SNR for these signals. Doing so will inevitably make use of the numerical simulation data at hand; we therefore stress that this calculation is only meant as a back-of-the-envelop estimate. Keeping this caveat in mind, we estimate that a post-merger SNR of 5 can be obtained for a system at $\sim 20$Mpc, assuming the DD2 EoS. Reference~\cite{2013PhRvD..88d4042R} estimates that the NS radius can be measured to $\sim 1.3$km at a distance of $300$Mpc using information from the premerger signal. Measurement accuracy scales proportionally to the distance, so a radius measurement to within $85$m is expected at a distance of $20$Mpc, which is comparable to the post-merger bound obtained here. We stress that both the premerger and the post-merger estimate ignore systematic uncertainties in the waveform models and the $f_{\peak}/M-R_{1.6}$ relation respectively; this calculation is meant as a comparison of the statistical errors only.

\subsection{Signal Energy}
\label{sec:energy}
\begin{figure}[h!]
\includegraphics[width=\columnwidth,clip=true]{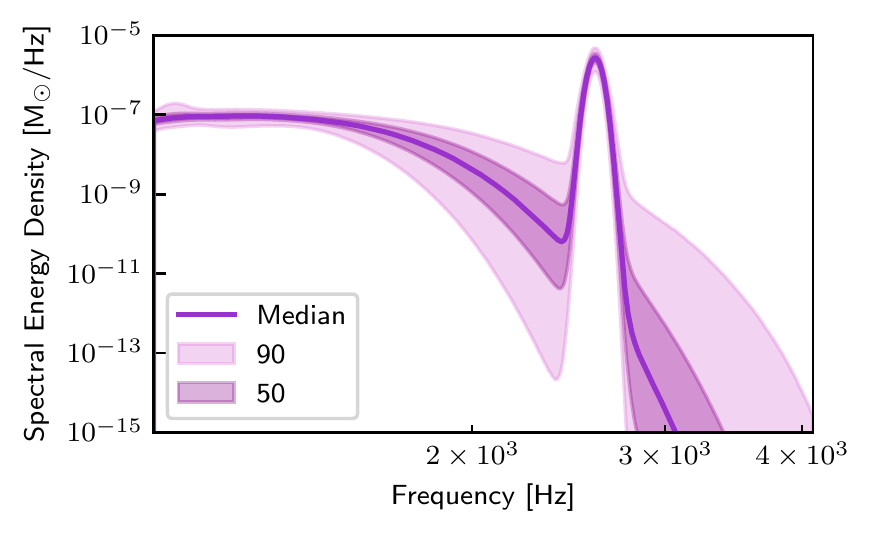}
 \caption{\label{fig:sedexample}SED posterior for the same system as Fig.~\ref{fig:dd2_135135_reconstruction} as a function of the frequency. The shaded regions denote $50\%$ and $90\%$ CIs of the posterior.}
\end{figure}
\begin{figure}[h!]
\includegraphics[width=0.9\columnwidth,clip=true]{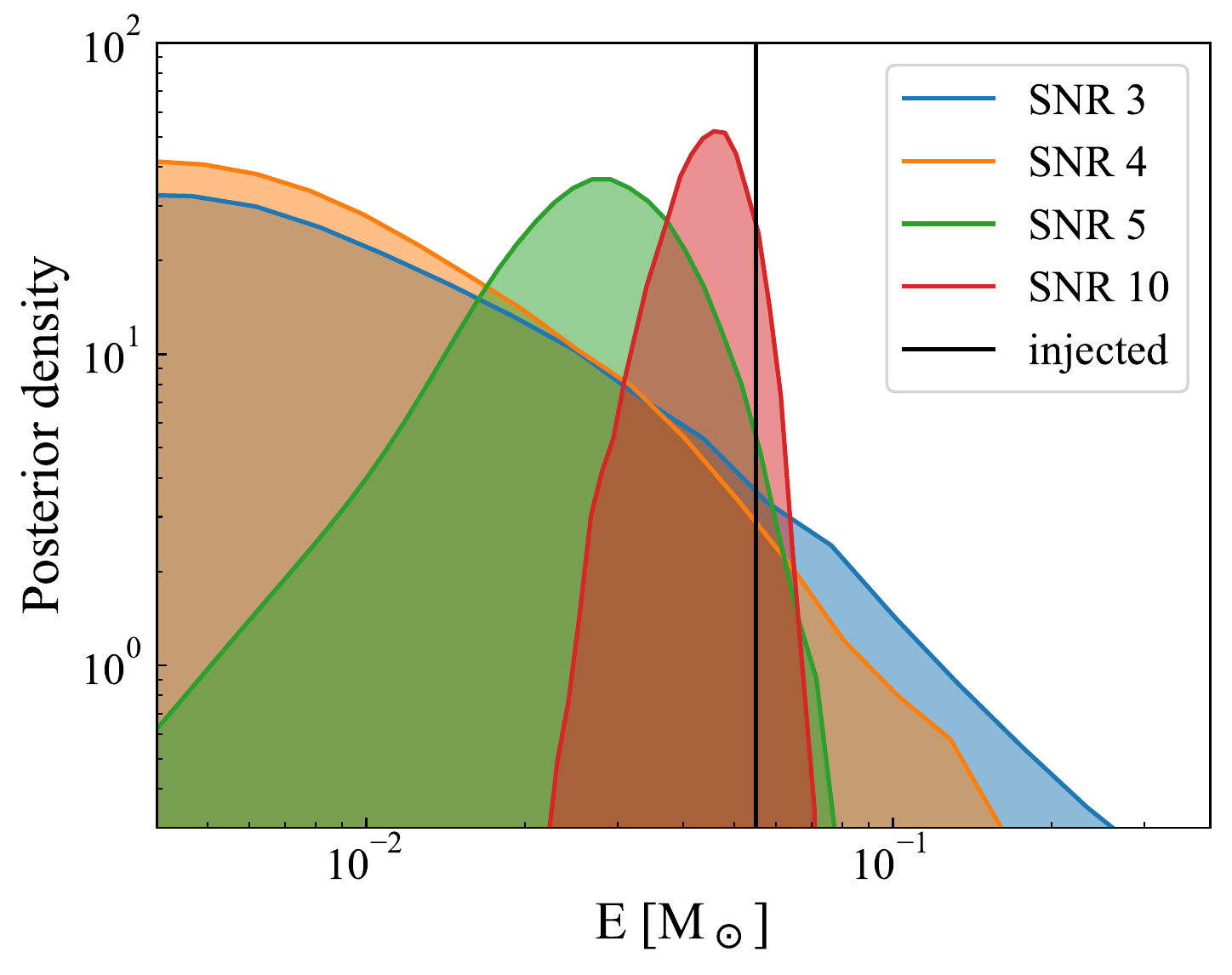}\\
\includegraphics[width=0.9\columnwidth,clip=true]{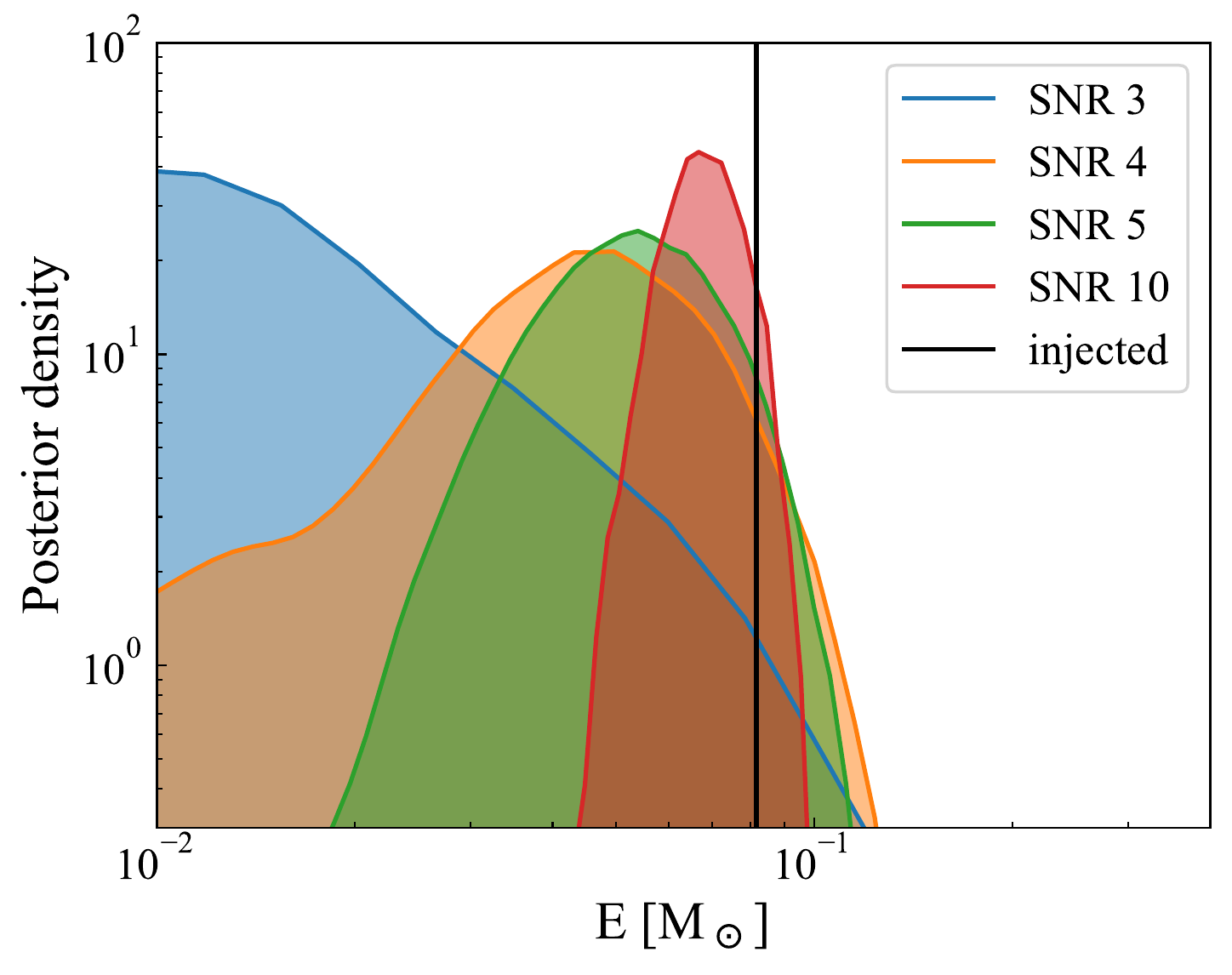}\\
\includegraphics[width=0.9\columnwidth,clip=true]{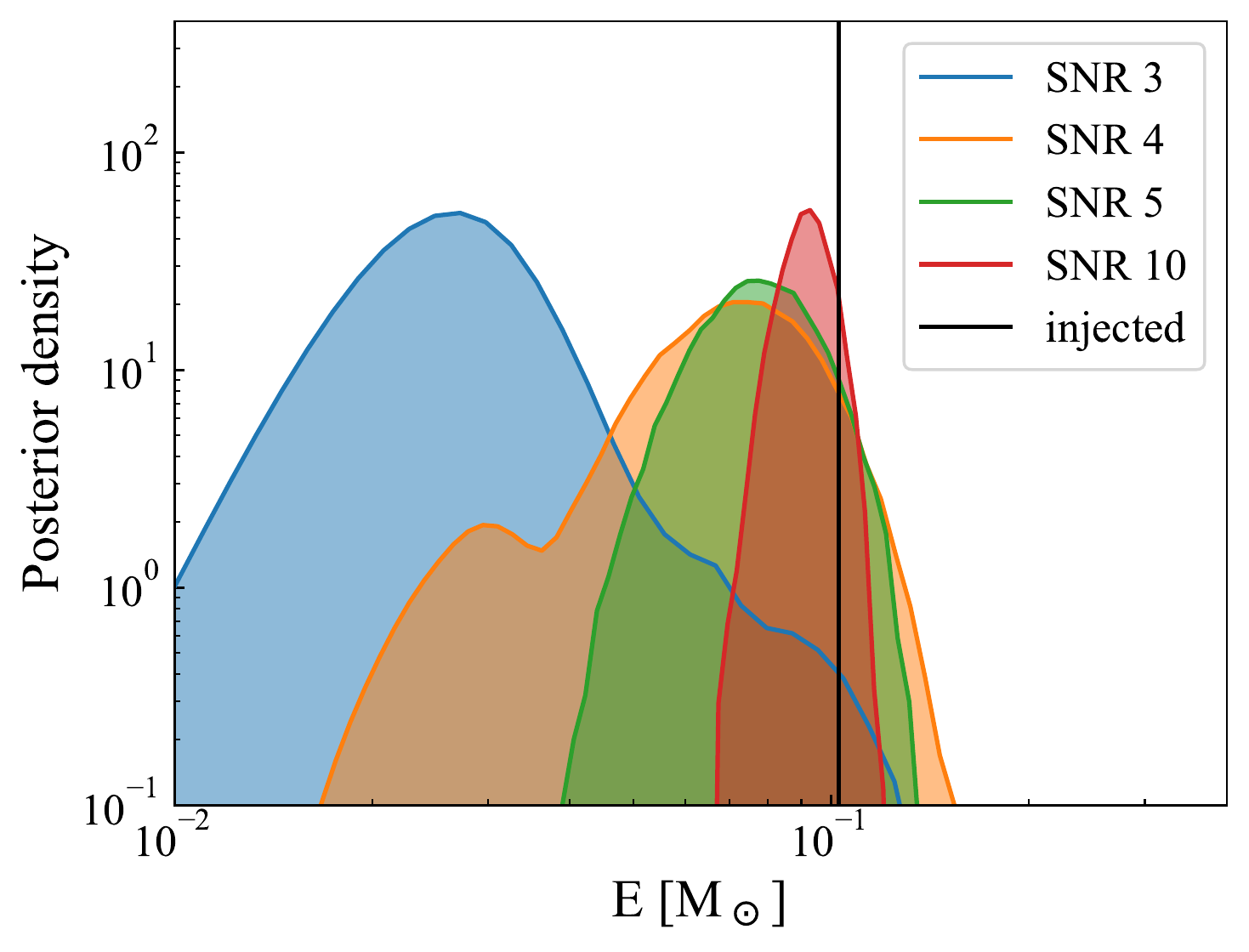}
\caption{\label{fig:energyposteriors} Energy  posterior density function for NL3 (top), DD2 (middle), and SFHO (bottom) for different post-merger SNR values. The vertical black line denotes the true injected energy. When the SNR is low and the signal is not reconstructed by {\tt BayesWave} the energy posterior can be used to place upper bounds on the energy emitted.}
\end{figure}
\begin{figure}[h!]
\includegraphics[width=0.9\columnwidth,clip=true]{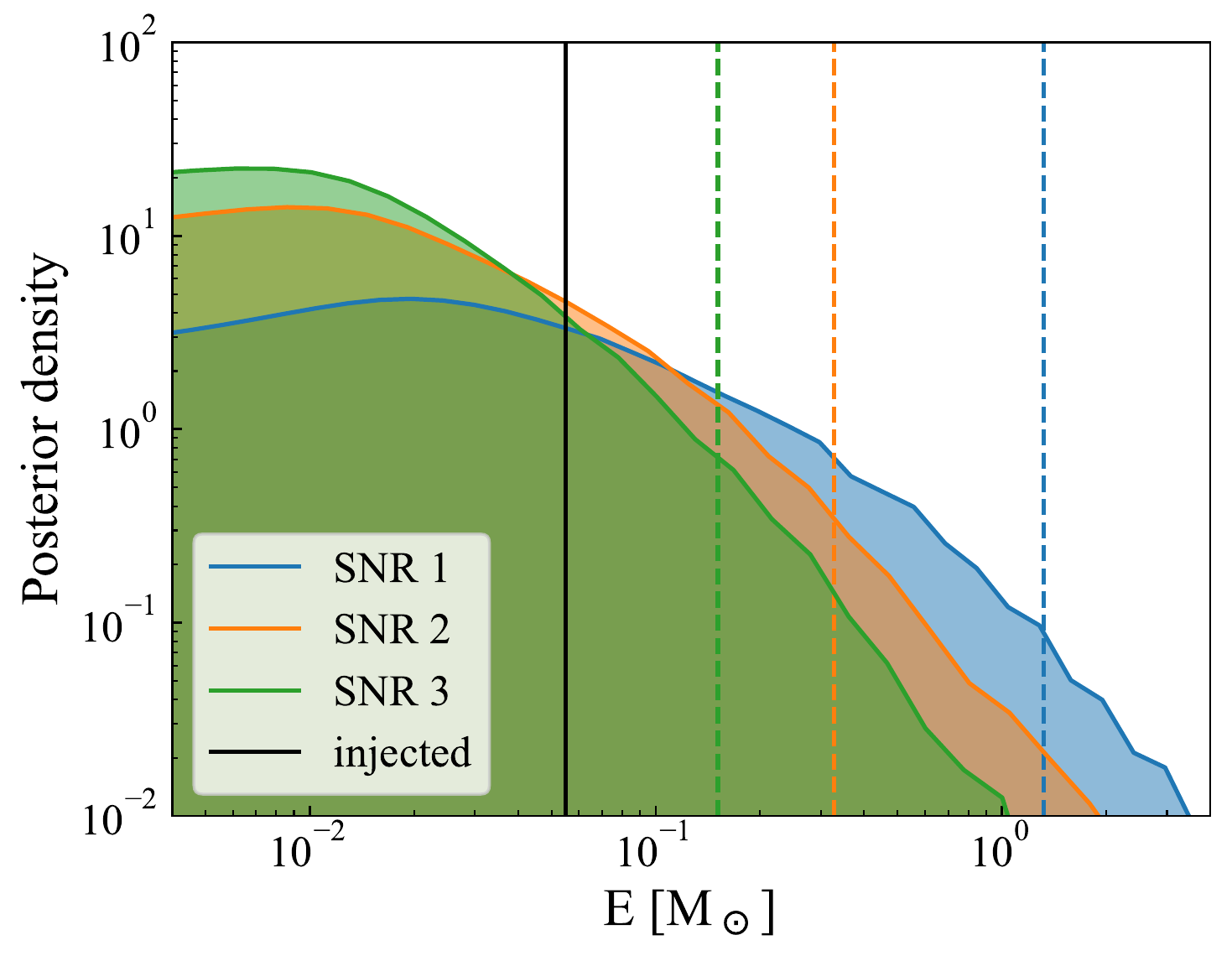}
\caption{\label{fig:energyposteriors_UL} Energy  posterior density function for NL3. We plot the energy posteriors for injections for which the signal was not reconstructed. The solid vertical line is the value of the injected energy, while the dotted vertical lines are the $95\%$ Bayesian UL obtained from from each injection. This UL can be used to place astrophysically interesting bounds on the energy emitted in the case of a nondetection of a post-merger signal from a confirmed BNS inspiral.}
\end{figure}

Besides quantities associated with the peak of the spectrum, the reconstructed spectrum can also be used to estimate the energy emitted in GWs and the \sed{}.  The GW flux is\footnote{Throughout this section we use units in which $G=c=1$.}
\begin{equation}
F_{\rm{GW}} = \frac{1}{16\pi} \langle \dot{h}_+^2(t) + \dot{h}_{\times}^2\rangle,
\end{equation}
where angle brackets indicate time averaging over the duration of the waveform and $h_+(t)$ and $h_{\times}$ are the plus and cross polarizations respectively. For a signal with effectively finite duration $\lesssim T$, the time-averaged flux is~\cite{SuttonBursts}
\begin{equation}
F_{\rm{GW}} = \frac{\pi }{4} \frac{1}{T} \int_{-\infty}^{\infty} \mathrm{d}f~f^2 \left(|\tilde{h}_+(f)|^2 + |\tilde{h}_{\times}(f)|^2 \right),
\end{equation}
and the total GW energy emitted is obtained by integrating over a sphere with a radius $D$, the distance to the source
\begin{equation}\label{eq:energy}
E_{\rm{GW}} = \frac{\pi}{4} D^2 \int \mathrm{d}\Omega \int_{-\infty}^{\infty} \mathrm{d}f~f^2 \left(|\tilde{h}_+(f)|^2 + |\tilde{h}_{\times}(f)|^2 \right).
\end{equation}
For BNS coalescences the GW emission is dominated by the $\ell=|m|=2$ mode so that the polarizations depend on the angle between the line of sight of the observer and the rotation axis $\iota$ as $h_+(t) \sim (1+\cos^2\iota)/2\, h_{\iota}(t)$ and $h_{\times}(t) \sim \cos \iota\, h_{\iota}(t)$.  Integrating Eq.~\ref{eq:energy} over the solid angle $\Omega$ gives
\begin{align}
E_{\rm{GW}} &=\frac{\pi}{4} D^2\int_{-1}^{1}  \mathrm{d} \cos{\iota}\left[  \frac{(1+\cos^2{\iota})^2}{4}+\cos^2\iota\right]\int_0^{2\pi}  \mathrm{d}\phi  \nn
\\
&\times \int_{-\infty}^{\infty} \mathrm{d}f~f^2 |\tilde{h}_{\iota}(f)|^2\nn
\\
& =  \frac{4}{5}\pi^2D^2  \int_{-\infty}^{\infty} \mathrm{d}f~f^2 |\tilde{h}_{\iota}(f)|^2,\label{eq:energyfinal}
\end{align}
where $\tilde{h}_{\iota}(f)$ is the Fourier transform of $h_{\iota}(t)$
and the \sed{} is
\begin{equation}
\frac{\mathrm{d}E_{\mathrm{GW}}}{\mathrm{d} f} = \frac{4}{5} \pi^2 D^2 f^2 |\tilde{h}_{\iota}(f)|^2.
\end{equation}
An example SED posterior is shown in Fig.~\ref{fig:sedexample}. We use the same injected system as for Fig.~\ref{fig:dd2_135135_reconstruction} and plot the median, $50\%$, and $90\%$ CIs. As expected, most of the energy is accumulated in the region of the spectrum peak. 

Figure~\ref{fig:energyposteriors} shows the posterior for the signal energy emitted in  $(1024,4096)$Hz for our three EoS at different injected post-merger SNRs. When the SNR is too low and {\tt BayesWave} does not reconstruct the injected signal (for example the SNR 3 case with NL3) the energy posterior peaks at low energy values. Despite not leading to definitive detection of post-merger emission, such a measurement could still be of astrophysical interest as it places an upper limit (UL) on the energy emitted. On the contrary, for high SNR signals, the post-merger signal is faithfully reconstructed and the energy posterior peaks more and more sharply at the expected injected value. Note, however, that {\tt BayesWave} tends to underestimate the median energy of the signal. This is because {\tt BayesWave} does not use an exact model for the signal but a decomposition in wavelets. This decomposition inevitably leads to imperfect signal reconstruction, as also demonstrated from the overlap not reaching the maximum value of 1 in Fig.~\ref{fig:overlapposteriors}. However, the injected value for $E_{\rm{GW}}$ is always included in the $90\%$ region of the full posterior, showing that we can still obtain a reliable estimate on the energy.

{\tt BayesWave}'s ability to provide astrophysically interesting and robust Bayesian ULs for the energy emitted is further demonstrated in Fig.~\ref{fig:energyposteriors_UL}. In this plot we show the energy posterior density for NL3 for three injections for which the signal was not reconstructed (overlaps consistent with $0$ in Fig.~\ref{fig:overlapposteriors}). The dotted vertical lines denote the $95\%$ Bayesian UL obtained from each injection. In the case of a nondetection of a post-merger signal following a known and detected BNS inspiral, this bound can provide an astrophysically interesting Bayesian UL on the energy emitted in the $(1024,4096)$Hz bandwidth.

\section{Monte Carlo Validation}
\label{sec:montecarlo}

%
\begin{figure}[h!]
\includegraphics[width=0.9\columnwidth,clip=true]{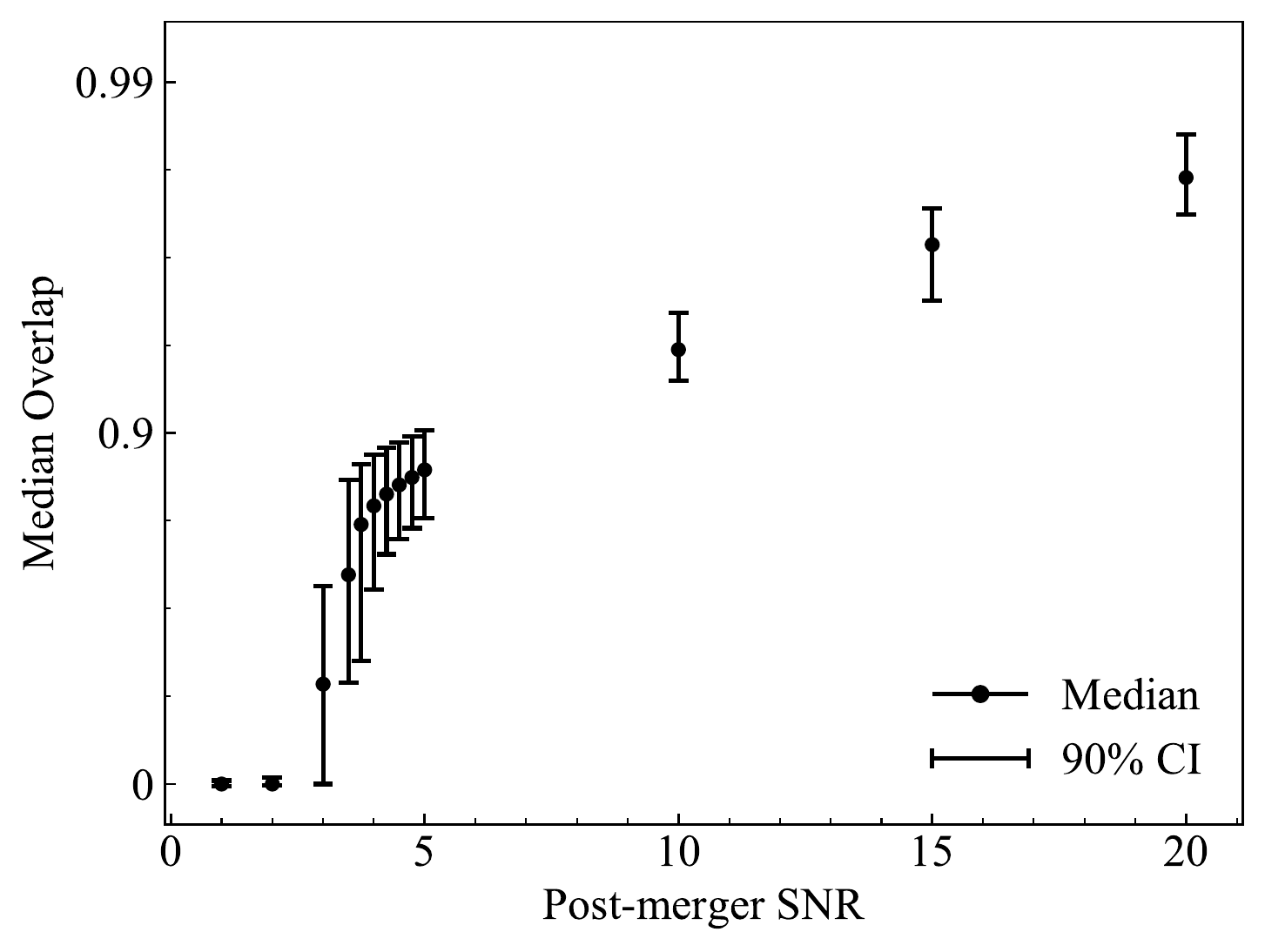}\\
\includegraphics[width=0.9\columnwidth,clip=true]{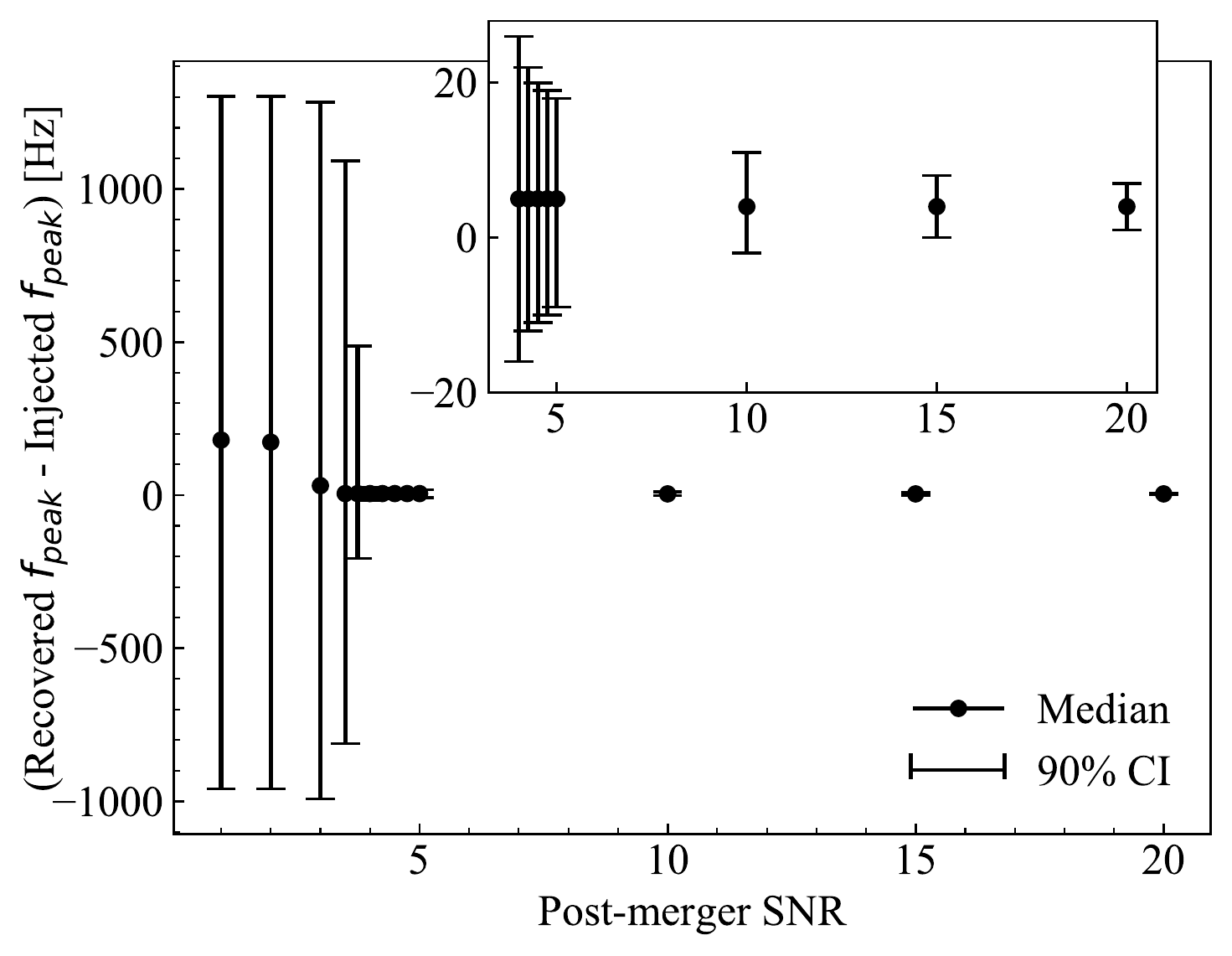}
 \caption{\label{fig:dd2_135135_Overlap-vs-SNR} Median over the population median and $90\%$ CIs for the overlap (left) and the error in the peak frequency (right) as a function of the SNR for 504 injections with DD2. The inset in the right panel demonstrates the peak frequency uncertainty at high SNR values for which the data are informative.}
\end{figure}
In the previous section we described in detail the full analysis of selected systems and discussed the reconstruction quality for different EoS and SNRs. In this section, we study statistical ensembles of systems in order to quantify the expected average results from a future BNS detection. Through Monte Carlo methods we create $504$ signals with DD2 and $m_1=m_2=1.35M_{\odot}$ with different SNRs and use {\tt BayesWave} to reconstruct their signal in a network of advanced detectors without a noise realization.

Figure~\ref{fig:dd2_135135_Overlap-vs-SNR} shows the median $90\%$ CI and median overlap (top) and error in the peak frequency (bottom) as a function of the SNR. As expected from the discussion of Sec.~\ref{sec:overlap}, the overlap values increase as the SNR increases. At low SNRs the signal reconstruction is not accurate, and the recovered overlaps cluster around zero. As the SNR increases, so do the overlap values, reaching $\sim 0.9$ at a post-merger SNR of 5.

Similar conclusions can be drawn from the bottom panel of Fig.~\ref{fig:dd2_135135_Overlap-vs-SNR}, where we plot the median over the 504 injections median and $90\%$ CIs for $f_{\peak}$. At low SNR values, {\tt BayesWave} does not reconstruct the signal; hence, the measurement is uninformative. At approximately SNR$\sim 4$, the signal becomes strong enough that the $f_{\peak}$ posterior starts deviating from the prior, achieving a measurement of $f_{\peak}$ to about $27$Hz at the $90\%$ level at a post-merger SNR of 5. This measurement accuracy is similar to the one obtained for the system extensively studied in Sec.~\ref{sec:reconstruction}.

\begin{figure}[h!]
\includegraphics[width=0.9\columnwidth,clip=true]{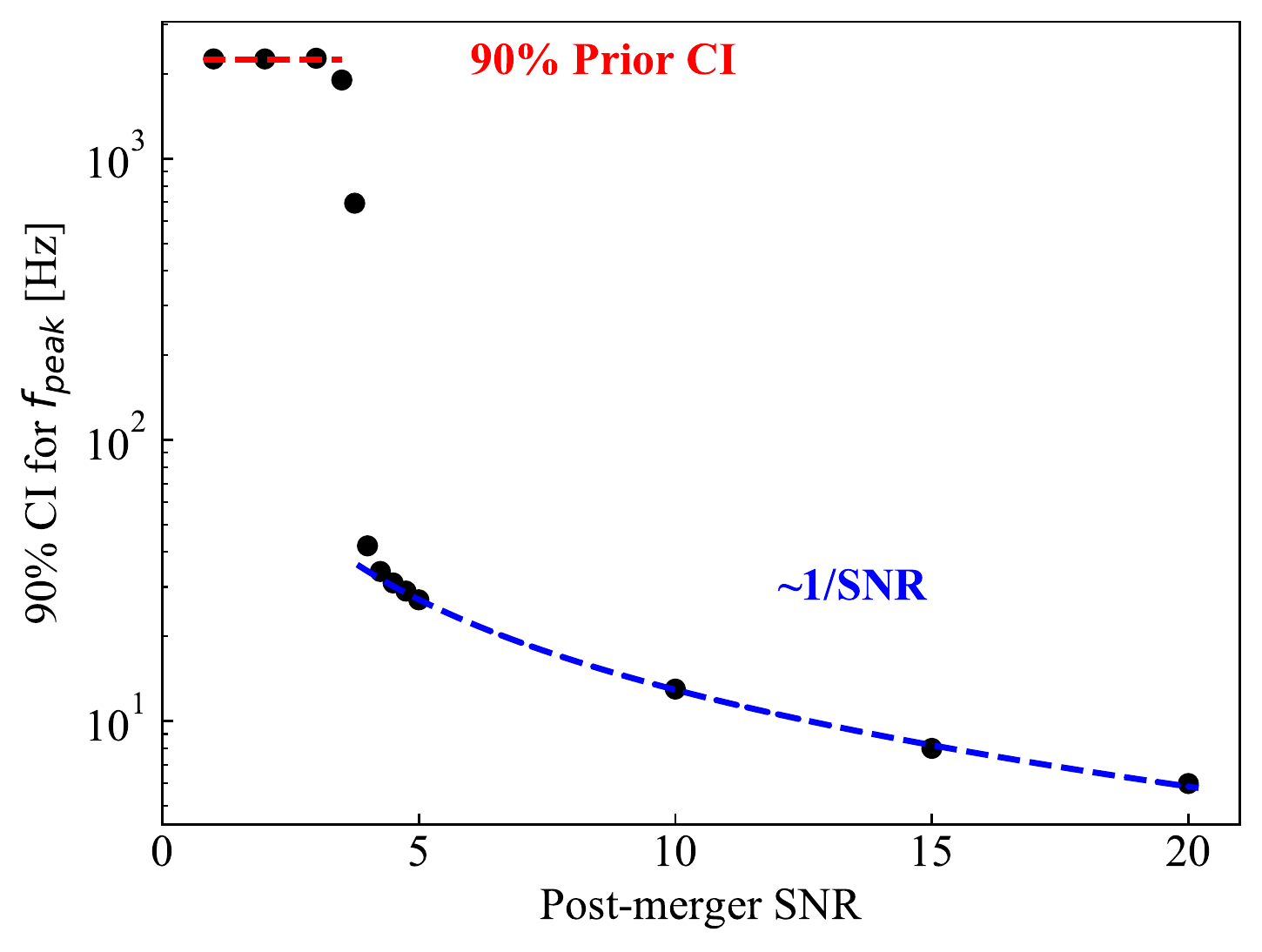}
 \caption{\label{fig:dd2_135135_deltafpeak-vs-SNR} Median $90\%$ CIs for the peak frequency as a function or the SNR. For SNR $\lesssim 4$, {\tt BayesWave} does not reconstruct the signal and the posterior CI is equal to the prior CI. For  SNR $\gtrsim 4$, {\tt BayesWave} reconstructs the signal and achieves the usual 1/SNR performance of matched-filtering analyses.}
\end{figure}

Figure~\ref{fig:dd2_135135_deltafpeak-vs-SNR} quantitatively studies the relation between the median $90\%$ CIs for the peak frequency and the SNR. For low values of SNR, {\tt BayesWave}  does not reconstruct the signal, and the $90\%$ posterior CI is equal to the $90\%$ prior CI. At high SNRs though, the width of the CI is proportional to 1/SNR, the expected scaling for matched-filter analyses.

We demonstrated in Sec.~\ref{sec:radius} that the systematic uncertainty of the $f_{\peak}/M-R_{1.6}$ universal relation is always larger than our statistical measurement error, assuming {\tt BayesWave} can reconstruct the signal. We therefore do not present a plot of the radius CI as a function of the SNR, but note that the error is around $500$m regardless of the SNR$\gtrsim 4$, and the error budget is dominated by the systematic uncertainty. i.e. the intrinsic scatter in the frequency-radius relation. 

\section{Conclusions}

We presented and studied a model-agnostic approach to extract information from the post-merger GW signal emitted during a BNS coalescence. Our method is fully generic, making only minimal assumptions about the underlying signal morphology. Despite this, we demonstrated that it is capable of reconstructing the post-merger signal. We described in detail how the reconstruction achieved can be used to measure the frequency of the peak of the post-merger spectrum. This measurement, in turn, can be used to place bounds on the NS radius by means of an existing EoS-independent universal relation. We showed that our analysis error is dominated by the intrinsic scatter in the universal relation, rather than the statistical error of the reconstruction. We leave detailed exploration of other existing universal relations for future work~\cite{2015arXiv150401764B,Lehner:2016lxy}.

We argued that information from the post-merger signal can lead to constraints on the NS EoS that are competitive with constraints originating from the premerger phase. However, the post-merger constraints studied here assume the existence of a loud-enough signal for {\tt BayesWave} to unambiguously detect. Even though it is unlikely that current ground-based based detectors will be fortunate enough to observe such a loud event, similar constraints can be achieved by combining information from a large number of dimmer signals~\cite{PhysRevLett.111.071101,Bose:2017jvk,Yang:2017xlf}. Detailed exploration of constraints obtainable from realistic populations of BNS coalescences are the subject of ongoing investigations. 

We stressed that our approach makes only minimal assumptions about the signal morphology and reduces systematic uncertainties. In parallel, \textit{BayesWave} has the flexibility to incorporate available information from BNS simulations in the form of Bayesian priors, should that information be deemed reliable. The more well-grounded prior information we can safely incorporate, the more sensitive the final analysis becomes. We plan to explore such targeted analyses that fall between general model-agnostic analyses and full matched filtering in the future. This approach will enable {\tt BayesWave} to more efficiently extract information about the EoS as well as analyze longer-duration signals.

As a final note we highlight our main result, namely that the statistical error in the NS radius measurement from the post-merger signal is comparable to the corresponding error from the premerger signal. We emphasize that these conclusions concern the statistical errors only. Future BNS simulations have to quantify the systematic uncertainties of the simulation data that form the basis for the empirical relations employed to invert frequency measurements to EoS properties. We anticipate that the intrinsic scatter in the empirical relations may be reduced by means of a better understanding of these relations including a physically motivated selection of candidate EoS.

\section{Acknowledgments}

We thank Carl-Johan Haster and Aaron Zimmerman for numerous engaging conversations. We thank Christopher P. L. Berry for sharing with us the fit for the total mass measurement error as a function of the SNR computed in~\cite{Farr:2015lna}. We thank Will Farr and Jonah Kanner for comments on the analysis and the manuscript. J.C. acknowledges support from NSF awards PHYS-1505824 and PHYS-1505524. A.B. acknowledges support by the Klaus Tschira Foundation.   M.M. and N.C. acknowledge support from NSF award PHY-1306702. This research was done using resources provided by the Open Science Grid~\cite{pordes:2007,Sfiligoi:2009}, which is supported by the National Science Foundation award 1148698, and the U.S. Department of Energy's Office of Science.  This research was also supported in part through research cyberinfrastructure resources and services provided by the Partnership for an Advanced Computing Environment (PACE) at the Georgia Institute of Technology~\cite{PACE}. Figures in this manuscript were produced using {\tt matplotlib}~\cite{Hunter:2007}.

\bibliography{refs}

\end{document}